\documentclass[preprint,3p,times,authoryear]{elsarticle}

\usepackage{graphicx}
\usepackage{amsmath}
\usepackage{amssymb}
\usepackage{amsfonts}
\usepackage{mathrsfs}
\usepackage{adjustbox}
\usepackage{mdframed}
\usepackage[figuresright]{rotating}

\usepackage[final]{changes}

\usepackage{lipsum}

\usepackage{dutchcal}

\graphicspath{ {figures/} }

\usepackage{nomencl}
\usepackage{etoolbox}
\usepackage{wrapfig}
\usepackage{framed} % Framing content
\usepackage[skins,breakable, most]{tcolorbox}

\renewcommand*\nompreamble{\begin{multicols}{2}}
\usepackage{textgreek}

\usepackage{epsfig}
\usepackage{epstopdf} %transfers eps into pdf before preparation
\usepackage{todonotes}
\usepackage{setspace}
\usepackage{siunitx}
\usepackage{tabularx}
\usepackage{longtable} 

\usepackage[utf8x]{inputenc} %Umlaute
\usepackage{stfloats}
\usepackage{verbatim}

\usepackage[caption=false]{subfig}
\usepackage[hyphens]{url}
\usepackage{graphicx}
\usepackage{booktabs}

\usepackage{lmodern,textcomp}
\usepackage{rotating} % Rotating table
\usepackage{multicol} % Multiple columns environment
\usepackage{multirow} 
\usepackage{mathtools, cuted}
\usepackage{lipsum}
\usepackage{tabularx} % allows defining width of tables
\newcolumntype{L}[1]{>{\raggedright\arraybackslash}p{#1}} % linksbündig mit Breitenangabe
\newcolumntype{C}[1]{>{\centering\arraybackslash}p{#1}} % zentriert mit Breitenangabe
\newcolumntype{R}[1]{>{\raggedleft\arraybackslash}p{#1}} % rechtsbündig mit Breitenangabe
\usepackage{amssymb}
\usepackage{pifont}% http://ctan.org/pkg/pifont
\RequirePackage[ngerman=ngerman-x-latest]{hyphsubst}
\usepackage{microtype}
\usepackage[american]{babel} 
\usepackage{hyperref}
\usepackage{float}
\usepackage[normalem]{ulem}
\usepackage{colortbl}
\useunder{\uline}{\ul}{}
\usepackage{lscape}
\usepackage{marvosym}

\usepackage{amsmath, amssymb}

\usepackage{lineno}
\biboptions{square,comma,sort&compress}

\setcitestyle{numbers}
\journal{}
\makeatletter
\def\ps@pprintTitle{%
 \let\@oddhead\@empty
 \let\@evenhead\@empty
 \def\@oddfoot{}%
 \let\@evenfoot\@oddfoot}
\makeatother
\usepackage{longtable}  %for better-spaced tables
\usepackage{tabulary}

\renewcommand{\nompreamble}{This list presents relevant symbols that are used within the body of this work. Most scalars will be denoted with cursive letters (mathcal style), function with gothic letters (mathfrak style), indices, graphs and most sets with roman letters and systems with cursive letters (mathscr style)}

\renewcommand\nomgroup[1]{%
  \item[\bfseries
  \ifstrequal{#1}{S}{Sets}{%
  \ifstrequal{#1}{F}{Functions}{%
  \ifstrequal{#1}{I}{Indices}{%
  \ifstrequal{#1}{M}{Vectors \& Matrices}{%
  \ifstrequal{#1}{G}{Graphs}{%
  \ifstrequal{#1}{Y}{Systems}{%
  \ifstrequal{#1}{C}{Scalars}{}}}}}}}%
]}
\makeglossary
\makenomenclature

\begin{document}

\begin{frontmatter}


\title{Active peer effects in residential photovoltaic adoption: evidence on impact drivers among potential and current adopters in Germany}


%\author[DTU,IIRM]{Author\corref{cor1}}
\author[DTU,IIRM]{Fabian Scheller\corref{cor1}}
\ead{fjosc@dtu.dk}
\cortext[cor1]{Corresponding author}
\author[IIRM]{Sören Graupner}
\author[SINUS]{James Edwards}
\author[KIT]{Weinand Jann}
\author[IIRM]{Thomas Bruckner}

\address[DTU]{Energy Economics and System Analysis, Division of Sustainability, Technical University of Denmark (DTU), Denmark}
\address[IIRM]{Institute for Infrastructure and Resources Management (IIRM), Leipzig University, Germany}
\address[SINUS]{SINUS Markt- und Sozialforschung GmbH, Berlin, Germany}
\address[KIT]{Institute for Industrial Production, Karlsruhe Institute of Technology (KIT), Germany}

\begin{abstract}
While the importance of peer influences has been demonstrated in several studies, little is known about the underlying mechanisms of active peer effects in residential photovoltaic (PV) diffusion. Empirical evidence indicates that the impacts of inter-subjective exchanges are dependent on the subjective mutual evaluation of the interlocutors. This paper aims to quantify, how subjective evaluations of peers affect peer effects across different stages of PV adoption decision-making. The findings of a survey among potential and current adopters in Germany (N=1,165) confirm two hypotheses. First, peer effects play a role in residential PV adoption: the number of peer adopters in the decision-maker’s social circle has a positive effect on the decision-maker’s belief that their social network supports PV adoption; their ascription of credibility on PV-related topics to their peers; and their interest in actively seeking information from their peers in all decision-making stages. Second, there is a correlation between the perceived positive attributes of a given peer and the reported influence of said peer within the decision-making process, suggesting that decision-makers’ subjective evaluations of peers play an important role in active peer effects. Decision-makers are significantly more likely to engage in and be influenced by interactions with peers who they perceive as competent, trustworthy, and likeable. In contrast, attributes such as physical closeness and availability have a less significant effect. From a policymaking perspective, this study suggests that the density and quality of peer connections empower potential adopters. Accordingly, peer consultation and community-led outreach initiatives should be promoted to accelerate residential PV adoption.
\end{abstract}

\begin{keyword}
Peer effects \sep Communication behaviour \sep Solar photovoltaic \sep Decision-making  \sep  Quantitative survey \end{keyword}

\end{frontmatter}
\section*{Highlights}
\begin{itemize}
\item Active peer effects are investigated and quantified across the PV adoption decision-making process.
\item Peer adopters stimulate PV decision-makers' interest in actively informing themselves about PV.
\item Effective exchanges about PV are dependent on the quality of peer-to-peer connections.
\item PV decision-makers are more likely to engage with peers to whom they ascribe positive attributes.
%\item PV decision-makers are more likely to engage with and more likely to be influenced by peers to whom they ascribe positive attributes.
\item Higher influence of credible peers than peers who are physically nearby, throughout all decision stages.
%\item Credible peers have more impact on PV decision-makers than peers who are merely physically nearby or accessible, throughout all decision stages.
%\item PV decision-makers who benefit from a dense network of credible and trustworthy peers are more likely to progress through successive decision stages and are more likely to adopt PV.
%\item Emotionally close peers in spatial proximity exhibit major influence

\end{itemize}

%\linenumbers

\newpage
\section{Introduction}
\label{S:1}

\subsection{Active peer effects in residential photovoltaic diffusion}
Reducing greenhouse gas emissions requires advances in all parts of the economy and society. Individual households can contribute to the required socio-economic transformation by adopting low-carbon consumption patterns \cite{Lin.2015}. On an individual household level as well as an infrastructural level, importance should thus be attached to the adoption of technical innovations that reduce carbon emissions. By reducing emissions, low-carbon innovations for households help mitigate anthropogenic climate change and thus contribute to the public good \cite{Sierzchula.2014}. This makes the diffusion of such innovations across society desirable from a societal point of view, and also explains why their adoption is currently pursued as a public policy goal. However, top-down strategies to accelerate the societal uptake of such innovations have not yet been successful on a sufficiently large scale \cite{Lin.2012, Clausen.2011, Geels.2018, Huber.2011}. On the one hand, the requisite technologies are insufficiently attractive from a consumer point of view: they tend to be perceived as expensive, poorly performant, and risky or uncertain due to their relative novelty \cite{Geels.2018}. On the other hand, research indicates that often it is neither the perceived technical feasibility of low-carbon innovations nor their perceived economic viability, that holds back their widespread diffusion. Rather, a range of cultural, institutional, and social barriers appear to be responsible \cite{Barnes.2019}.

While the residential uptake of PV systems has been approached from various angles \cite{Scheller.2021,Rode.2020,Rai.2016b}, the importance of peer influence in shaping residential adoption decisions has been recognized by a number of studies. According to Wolske et al.~\cite{Wolske.2020}, within the energy context, the term “peer effects" encompasses the impact of peer attitudes, values, behaviours, and interactions on the attitudes, values, or behaviours of individual decision-makers. In other words, “peer effects are the influence of a persons' peers [...] on his or her behaviour”\cite{Palm.2017}. Across the literature, peer groups are often defined in terms of spatial or geographic proximity: for instance, all private households within a neighbourhood or a district can be defined as “peers" \cite{Bollinger.2012,Richter.2013,Graziano.2015,Rode.2020}. Such definitions implicitly assume an objective or etic (outsider) viewpoint. However, it is quite possible that residential decision-makers themselves might not consider all spatially proximate households to be “peers" \cite{Wolske.2020}. From a subjective or emic (insider) viewpoint, “peers" might be defined as members of a decision-maker's social network or individuals who share their interests or values, regardless of spatial proximity \cite{Rai.2016b,Palm.2016,Palm.2017,Wolske.2017}. “Peers" as defined subjectively, based on shared social networks and/or shared interests or values, are more likely to be connected with and known by one another than “peers" as defined based on geographical proximity alone \cite{Wolske.2020}. For the rest of this paper, “peers" are defined from a subjective viewpoint as people who are connected with, known by, or regarded as “similar" by residential decision-makers themselves. This comprises family members, relatives, friends, acquaintances, and co-workers.

Peer effects can be further distinguished into two components: Active processes and passive processes. According to Bollinger and Gillingham~\cite{Bollinger.2012}, peer effects on potential PV adopters develop as a consequence of the visibility of the PV panels themselves (passive peer effects \cite{Rai.2013}) and word-of-mouth, i.e. conversations about PV (active peer effects \cite{Rai.2013}). Passive peer effects are often assumed by researchers to represent a substantial part of the entire peer effect, as the visibility of PV systems is generally high \cite{Bollinger.2012, Rai.2013, Rode.2020}. However, decision-makers interviewed qualitatively by Palm \cite{Palm.2017} assessed active peer effects as more important than passive ones. 
%The author suspects, however, that it is still possible that the influence of seeing PV systems was highly underestimated by the decision-makers as they had difficulties in measuring the passive peer effect and they already had a high level of awareness of the technology. 
Rai and Robinson \cite{Rai.2013} assume that active peer effects may arise from passive peer effects, whereas Palm \cite{Palm.2017} could not find qualitative evidence that seeing PV systems fostered contacts between decision-makers and previous adopters who they did not already know personally. It is clear that research remains to be done on active peer effects, their relationship to passive peer effects, and the relative contribution of both to the entire peer effect. Research also remains to be done on the social and personal factors in play during active peer interactions: Wolske, for instance, assumes that the quality of a given peer connection is more important than geographical proximity or the peer \cite{Wolske.2017,Wolske.2020}.

Despite the importance of active peer effects in accelerating residential PV adoption, there is only a little knowledge about underlying social mechanisms \cite{Palm.2017,Rai.2013,Wolske.2020}. In particular, there is a significant gap in empirical work on how the impact of active peer effects is mediated through the decision-maker's perception of the peer \cite{Palm.2017,Rai.2016b}. Additionally, there is an indication that the social influence changes throughout the PV adoption process \cite{Scheller.2020,Scheller.2021,Arts.2011}.  
Many studies, however, hardly provide details on whether and how peer effects relate to different stages \cite{Palm.2017}. Addressing these gaps could provide insights for peer empowering more effectively and for improving the effectiveness of policy measures.

\subsection{Research objectives and contributions}

The aim of the present study is to explore how social peer influence affects residential PV adoption decisions throughout the decision-making process. Special attention is given to the contribution of perceived peer attributes to the mechanisms underlying active peer effects. The following research questions are used as a guideline: 

\begin{itemize}
\item What determinants stimulate residential decision-makers to adopt PV, and to what extent does the presence of peer PV adopters in decision-makers' social circles affect their PV-related beliefs and intentions? 
%\item With regard to PV, are decision-makers more likely to interact with peers to whom they ascribe positive attributes?
%\item With regard to PV, are decision-makers more likely to be influenced by interactions with peers to whom they ascribe positive attributes? 
\item What ascribed peer attributes have the most impact on the likelihood of active initiation of peer interaction and influence by such peer interactions throughout different stages in the residential PV decision-making process? 
%\item Do these effects differ at different stages in the residential PV decision-making process?
\end{itemize}

To answer these research questions, this paper uses a quantitative approach. A survey was conducted with potential and actual residential PV adopters living in owner-occupied homes in multiple regions of Germany. The survey questionnaire focused on residential decision-makers' active interactions with various peer groups on the topic of PV; their perceptions of various peer groups' attributes; and their assessments of these peer groups' influence on their PV decision-making process. Passive peer effects and overall attitudes toward PV were also measured, as were decision-makers' positions in the PV decision-making process. This enabled the identification of peer attributes that enhance the influence of active peer effects at different stages in the decision-making process. The survey questionnaire design drew upon the following theories: 1) the theory of innovation adoption provides a conceptual basis for modelling the residential decision-making process \cite{Rogers.2003,Wilson.2018}; 2) source credibility theory provides valuable insights on the impact of peers' perceived credibility on their ability to influence decisions \cite{nesler1993effect,ohanian1990construction}. This paper seeks to contribute to the basic understanding of peer effects in energy decision-making, as well as to the growing body of literature on policy strategies to speed the diffusion of low-carbon technologies.

The rest of this paper is structured as follows: Section~\ref{S:2} reviews the related work. Section~\ref{S:3} provides an overview of the survey design and the sample. Section~\ref{S:4} presents and analyzes the survey findings. Section \ref{S:5} discusses situates the findings in a theoretical context and derives practical implications. Finally, Section \ref{S:6} concludes with a summary and suggested directions for future work.

\section{Related work}
\label{S:2} 

\subsection{Impact drivers of active peer influences}
\label{sec:activepeer}
Research to date has demonstrated that active peer influences play a role in the PV decision-making process, with decision-makers often going out of their way to establish contact with PV-owning peers. Mundaca et al. \cite{Mundaca.2020} and Petrovich et al.~\cite{Petrovich.2019} found that the presence of PV systems in a decision-maker's local environment (neighbours) and personal social circle (friends and family) drive their likelihood of adopting to a significant degree. Among the decision-makers studied by Palm \cite{Palm.2017}, the strongest active peer effects resulted from personal contacts such as friends, acquaintances, or relatives, leading Palm to conclude that “established social connections [are] more important than geographical proximity”. Geographically proximate peers categorized by decision-makers merely as “neighbours", without any further personal relationship, were rarely perceived as influential, especially in comparison to local peers with whom decision-makers had some kind of personal relationship \cite{Palm.2016}. This is in line with social network analyses, which tend to show that the persuasiveness of word-of-mouth communications depends on the quality of the relationship between the interlocutors \cite{Wolske.2020}. Trust clearly plays a role here: Wolske et al. \cite{Wolske.2017} found a positive correlation between decision-makers' trust in their social networks, their interest in learning about PV systems owned by their peers, and their belief that theirs social network support PV adoption. This may be because insight shared by trusted personal contacts can reassure decision-makers of their competence to make an informed decision: Palm interprets such insight as “a confirmation from a trustworthy source (e.g., a person that the respondents knew and that was in a similar situation as themselves) that the technology worked as intended and without hassle” \cite{Palm.2017}. Rode and Müller’s \cite{Rode.2020} findings support Palm's conclusion as to the importance of personal contacts \cite{Palm.2017}. They additionally hypothesise that neighbours are more likely to have close personal relationships in rural areas than in urban areas, and therefore more likely to communicate with each other \cite{Rode.2020}. Thereby, peer effects had less of an effect with a higher diffusion rate \cite{Graziano.2015,Rode.2020}.

Contrary evidence appears in Rai et al. \cite{Rai.2016b}, who found that less than a fifth of the respondents mentioned “conversation with friend/family/work” as a primary spark event for PV adoption, and that the median response regarding the importance of information from family, acquaintances, and co-workers was rather low \cite{Rai.2016b}. The strength of personal relationships between a decision-maker and their PV-owning peers is clearly not the only factor influencing their decision; indeed, interactions between factors likely come into play. Mundaca and Samahita \cite{Mundaca.2020}, for instance, showed that while learning about PV systems from personal contacts increased decision-makers' likelihood of adoption, even stronger peer effects arose from interactions with peers identified as both geographically proximate and likeable. However, Palm \cite{Palm.2017} added that these results may be different due to contextual factors as “Swedes might be less prone to talk to their neighbours than are people in the U.S. or Germany”. Nevertheless, it can be assumed that likely related trustworthiness is relevant for active peer effects even if this might be of less importance in Germany. We can conclude that a better understanding of the impact of active peer effects would seem to require a more comprehensive evaluation of the various social, psychological, geographical, and other factors in play.

\subsection{Importance of peer credibility}
Considering that active peer effects are based on information gathered through different kinds of actively-triggered social influences, Berlo et al. \cite{berlo1969dimensions} point to the fact that “an individual’s acceptance of information and ideas is based in part on ‘who said it’”. Wolske et al. \cite{Wolske.2020} add that the persuasiveness of a word-of-mouth exchange depends on the “quality" of the connection between the interlocutors. Within the context of PV adoption, decision-makers appear significantly more likely to be influenced by interlocutors to whom they ascribe certain positive attributes, summarised here under the heading of credibility \cite{Scheller.2020}. 

Several authors have operationalized credibility in different ways \cite{ohanian1990construction}. Here, multiple dimensions of credibility are posited. Firstly, empirical evidence indicates that a message is perceived as most credible when the communicator is regarded as both an expert and trustworthy~\cite{mcginnies1980better}. Expertise is defined as having “adequate knowledge, experience or skills”~\cite{van2009celebrity}. Trustworthiness is related to “the degree of confidence in, and level of acceptance of, the [peer] and the message”~\cite{ohanian1990construction}, as well as the communicator's “honesty, integrity and believability”~\cite{van2009celebrity}. The findings of Lui and Standing~\cite{lui1989communicator} and McGinnies and Ward~\cite{mcginnies1980better} suggest that trustworthiness is at least as important for effective communication as expertise. A third factor that clearly bears upon the persuasiveness of messages is power, defined here as the perceived legitimate right of the communicator to influence the decision-maker and obtain their compliance \cite{steiner1965current}. Power and credibility are often intertwined \cite{nesler1993effect}.  %The attribute likeability is considered in addition to the most frequently mentioned dimensions expertness and trustworthiness.
Early research on innovation diffusion furthermore suggests that geographical proximity conditions the likelihood of social influence~\cite{hagerstrand1965monte}: the closer (and more available) the communicator, the higher the likelihood and strength of influence \cite{meyners2017role}. Finally, in addition to rational and objective factors, affective factors come into play. Specifically, communication tends to be more effective when the communicator is perceived as likeable; likeability, defined as “affective intimacy between source and its~receiver”~\cite{lui1989communicator}, might indeed be seen as a mediator between various other attributes.

\subsection{Procedural decision-making}
\label{S:2.1}
An innovation decision process or adoption process “examines the individual and the choices an individual makes to accept or reject a particular innovation”~\cite{Straub.2009}. It is analytically helpful to break such processes down into stages. Rogers’~\cite{Rogers.2003} adoption model, for instance, consists of five sequential stages. Similarly, Wilson et al. \cite{Wilson.2018} propose a three-stage process for home refurbishments. Various events, such as stakeholder interactions, can “spark" movement from one stage to another, for instance from the stage of awareness of PV to consideration or adoption ~\cite{Rai.2016b}. While stage models have been criticised for being overly linear and excluding contextual variation ~\cite{Wilson.2007, Arts.2011, Axsen.2012}, the literature has consistently indicated their utility\cite{Wolske.2017,Sun.2020,Parkins.2018}.

Accordingly, the present study builds upon an abstract four-stage process model for residential PV decision-making and adoption, the stages of which are awareness, interest, planning, and implementation \cite{Scheller.2020,Scheller.2021}. The decision-making process consists of the first three stages, while implementation completes the adoption process. The four-stage model takes account of the difference between simple awareness of the existence of a low-carbon technology and active decision-making. It furthermore provides a framework within which to analyse the way decision-makers accumulate information through interactions, building more object-specific knowledge over time. A given decision-maker's progression through the stages can consist of a sequence of increasingly concrete planning processes that demonstrate deeper engagement and higher likelihood of adoption.

\section{Investigation design}
\label{S:3}

\subsection{Form of data acquisition and design of the survey}
\label{S:3.1}
%The data acquisition on the communication behaviour was conducted only once at a specific time. Therefore, this investigation represents a cross-sectional study. 
Since this investigation aims to quantify perceptions representative of the population of current and potential residential PV decision-makers in Germany, a non-experimental design was chosen. %An exploratory sequential mixed methods design was chosen: first, qualitative focus group discussions were conducted with PV decision-makers and stakeholders in order to refine the research questions; 
A structured survey was conducted with PV decision-makers in order to ensure nationwide reach. Due to near-universal internet penetration within the target population of house owners with at least some affinity for advanced technology, an online survey approach (computer-assisted web interviewing or CAWI) was chosen as the most efficient means of realising the quantitative phase.

The German-language survey questionnaire covered question areas regarding PV awareness, interest, and intention to adopt; stage in the PV decision-making and adoption process; adoption drivers and barriers; presence of adopters in the decision-maker's geographical area and peer group; perception of the peer and stakeholder attributes; contact with peers; assessment of the influence of peer contacts; and decision-maker characteristics (socio-demographic attributes and social milieu). An overview of relevant terms are outlined in Table \ref{tab:Terms}. The questions themselves are provided in English translation in the Supplementary Material A. The survey was designed in collaboration with the institute SINUS Markt- und Sozialforschung GmbH\footnote{SINUS (https://www.sinus-institut.de/en/) is an independent institute operating in the field of social science research and consultancy. The institute offers a broad service portfolio for qualitative and quantitative research and is specialized in target group segmentation (Sinus-Milieus\textsuperscript{\textregistered}).}. %The segmentation is based on a social milieu approach that groups milieus which could be described loosely as groups of like-minded people, using social status, lifestyles and basic values. This approach is a powerful means for moving beyond the traditional sociodemographic criteria.. 

In order to ensure a tight focus on the target population of current and potential residential PV decision-makers, the survey questionnaire opened with a screener battery. Respondents who indicated that they owned a house capable of hosting a PV system (i.e. a single-family home or duplex) and were solely or jointly responsible for major technological or renovation investments were invited to proceed through the rest of the survey. These respondents were then sorted into two groups: those who had already acquired a PV system (current adopters, CAs) and those with a moderate or strong intention of purchasing a PV system (potential adopters, PAs). PAs were asked to indicate their perceived stage in the adoption decision process using a slider (100-point scale) matched to a graphic showing and describing the four stages: I) awareness stage; II) interest stage; III) planning stage; IV) utilization (adoption) stage. The stages were described as in \ref{tab:Terms}. CAs were automatically placed in IV) utilization stage.

\begin{table}
\footnotesize
\renewcommand{\arraystretch}{1.2}
\setlength{\tabcolsep}{5pt}
    \caption{Definition of the terms as used and presented in the survey. The first section of the table describes the different stages of decision-making. The middle and the end sections describe the relevant peer and stakeholder groups and peer group attributes, respectively. The survey respondents represented the group of residential PV decision-makers (current and potential adopters).}
    \label{tab:Terms}
\begin{tabularx}{\textwidth}{p{0.24\textwidth} p{0.75\textwidth}}
\toprule
\textbf{Term} & \textbf{Description} \\ 
\midrule
\multicolumn{2}{l}{\textbf{Decision-making process stages}}\\
\hline
Awareness & I am aware of the existence of residential PV systems (stage I)  \\
\hline
Interest & I find residential PV systems interesting and I want to know more about them (stage II)  \\
\hline
Planning & I plan to install a residential PV system in my home (stage III)  \\
\hline
Implementation & I have signed a contract for a residential PV system (decision process stage IV) \\
\midrule
\multicolumn{2}{l}{\textbf{Residential decision-makers}}\\
\hline
Potential adopters & Respondents who are interested in PV and find themselves in stage I-III\\
\hline
Current adopters & Respondents who have already acquired or signed a contract for a PV system and find themselves in stage IV\\
\midrule
\multicolumn{2}{l}{\textbf{Social peer groups}}\\
\hline
Family and relatives & Private persons among the decision-maker’s family and relatives  \\
\hline
Friends & Private persons within the decision-maker’s immediate social circle  \\
\hline
Acquaintances and co-workers & Private persons within the decision-maker’s wider social circle  \\
\hline
Neighbours & Private persons living in the decision-maker’s neighbourhood \\
\midrule
\multicolumn{2}{l}{\textbf{Peer group attributes}}\\
\hline
Trustworthiness & Peers offer honest and transparent statements and information about PV systems
  \\
\hline
Competence & Peers are very competent with regard to PV systems.
  \\
\hline
Power & Peers have the power to stand in the way of my decisions regarding PV systems \\
\hline
Independence & Peers would not profit from my decision to acquire a PV system
 \\\hline
Availability & Peers are always available for discussions or to provide information on PV systems
 \\\hline
Closeness & Peers are located near my place of residence and home
 \\\hline
Integrity & Peers behave in the way that I would generally expect
 \\\hline
likeability & Peers are likeable to me
 \\\hline
Reliability & Peers keep their promises and honour contracts
 \\
\midrule
\multicolumn{2}{l}{\textbf{Contact directions}}\\
\hline
Active initiations & I initiated contact with the peer group
  \\\hline
Passive experiences & The peer group initiated contact with me
  \\
\bottomrule
\end{tabularx}
\end{table}

In the following questionnaire section, “Ego", respondents were asked to provide a broader assessment of their relationship to PV technology. PAs were asked to indicate the strength of their intention to adopt residential PV, while CAs were asked to provide details about their residential PV acquisition and evaluate their PV systems (ten-point ordinal scale). Next, respondents were asked how strongly they agreed (five-point ordinal scale) with a range of statements covering relevant beliefs about adopting PV (i.e. rational and normative benefits) and relevant beliefs about behavioural control (i.e. financial, psychological, and contextual barriers). Tailored question phrasings were used for PAs and CAs.

The next questionnaire section, “Alteri", addressed passive and active peer effects. First, respondents were asked to indicate the approximate number of PV systems within their groups of neighbours, family, and friends, as defined in Table \ref{tab:Terms} (single choice per group: none; one to two; three to four; five or more). %(family and relatives, friends, acquaintances and co-workers, as well as neighbours). 
Second, respondents were asked to report their perception of the relevant attributes, as defined in Table \ref{tab:Terms} (sliding scale from 1=the selected attribute does not apply at all to the respective peer group to 10=the selected attribute applies completely to the respective peer group; additional option of “I don't know or can't judge"). %(trustworthiness, competence, power, independence, availability, closeness, likeability, integrity, reliability)
%of 14 peer groups: family and relatives; friends; acquaintences and colleagues; neighbours; other private persons with an interest in PV; local energy providers; state- or national-level energy providers; communal administrations; private organisations and societies; financial institutions; contractors, architects, and technical service providers; energy consultants; PV producers and providers; and reports/media 

Third, respondents were asked to provide information about the peer contacts that influenced them during each stage of the PV adoption process that they had completed, in reverse order (i.e. CAs and stage III PAs were first asked about contacts during stage III, then stage II, and finally stage I, while stage II PAs were asked about contacts during stage II, then stage I). Per relevant stage, respondents indicated which peer groups they had contact with; whether each contact was unidirectional or bidirectional (i.e. receiving information only vs. engaging in an exchange); whether each contact was initiated by the respondent or the peer; and whether each contact exerted influence on the respondent's PV adoption decision. For each contact to which respondents ascribed influence, the perceived strength of influence was then measured (sliding scale from 1=the selected peer group had very weak influence in the respective decision stage to 10=the selected peer group had a very strong influence in the respective decision stage). 

The final section of the survey comprised several personal questions used to compare different decision-making adopter segments: education level; income level; household status and composition; area of residence; perceived neighbourhood cohesion; and social milieu (Sinus-Milieus\textsuperscript{\textregistered}) (single-choice or multi-item scales, with options not to answer were appropriate). %Sinus-Milieus\textsuperscript{\textregistered} were assessed using a validated multi-item indicator provided by SINUS Markt- und Sozialforschung GmbH.

\subsection{Procedure of recruitment and sourced sample}
\label{S:3.3}

In line with the research objectives, data were obtained from single- or two-family house owners (residential decision-makers) who either own a PV system (sub-group CA) or are in advanced stages of the adoption decision process (sub-group PA). House owners who reported no interest in acquiring a PV system at all and/or who did not locate themselves in at least decision stage II were excluded from the survey. This decision was made in the knowledge of the fact that the results would be exploratory rather than representative. The sample was sourced from the Ipsos Online Access Panel\footnote{More information can be found here: https://www.ipsos.com/en/sample-access.} using a specialised sampling tool that makes a random selection of potential respondents. %The panel is used for social and market research only and the participants are recruited with the help of a multi-step process. 
Potential survey participants were invited to participate in the survey by personal e-mail. During the field time, the addressees received a maximum of two reminders to participate in the survey. The final sample consists of 1,165 completed questionnaires. Fieldwork was completed on Tuesday, December 17\textsuperscript{th} and the final dataset was delivered on Wednesday, December 18\textsuperscript{th} 2019. %This corresponds to a response rate of around 10\%. This is also attributable to the incentive for Ipsos panel participants. The incentive is based on a special point system, which is graded according to the length and complexity of the questionnaire. The points can either be donated to charity or redeemed by the participants in the form of vouchers. 

\begin{table}[h]
\footnotesize
\centering
\caption{Demographic structure of the total sample population (N) and the sample population with respect to the information concerning gender (Male; Female), net income ($\leq$ \EUR{2,340} low net income (Low); \EUR{2,341-3,200} lower-middle net income (L.-M.); \EUR{3,201-4,840} upper-middle net income (U.-M.); $\geq$ \EUR{4,841} high net income (High), and residence (City; Suburb; Country). The final sample consists of 1,165 respondents.}
 \label{tab:sample_overview}
\begin{tabular}{l|l|lll|ll l l| l l l }
\toprule
&  & \multicolumn{3} {  l  | }  {\textbf{Gender}} & \multicolumn{4} {l| }{\textbf{Income*}} & \multicolumn{3} {l}{\textbf{Residence}} \\ 
\textbf{Sub-group} & \textbf{N} & \textbf{Male} & \textbf{Female} & \textbf{Other} & \textbf{Low} & \textbf{L.-M.} & \textbf{U.-M.} & \textbf{High} & \textbf{City} & \textbf{Suburb} & \textbf{Country}  \\ 
\midrule
PA & 771    &        403      &  368 &  0  &  164 &    175 &     198  & 149 & 224 & 207 & 340 \\
CA & 394      &         227      &   166  &  1  &   48 &     67 &     116 & 118 & 169 & 97 & 128 \\
%            &      (36\%)   &  (31,1\%)      &      (48.4\%)   &  (55.6\%) &  (55.6\%)   \\
\midrule
Total  & 1,165            &     630   &     534 &  1   &     212 &     242 &    314 & 267 & 393 & 304 & 468  \\  
\bottomrule
\addlinespace[1ex]
\multicolumn{6}{l}{\scriptsize{\textsuperscript{*} study respondents without an indication are not shown}}
\end{tabular}
\end{table}

An overview of the sample in absolute numbers is outlined in Table \ref{tab:sample_overview}. Further insights into the relative sample compositions are presented in Figure \ref{fig:sample_compositions} and outlined in Supplementary Material B. The respondents are distributed between the sub-groups CAs (\textit{n=394}) and PAs (\textit{n=771}). Among PAs, around 75\% located themselves in stage II (\textit{n=602}), while the remainder located themselves in stage III (\textit{n=169}). A comparison of the total sample at hand with structural data on house owners in Germany shows that the sample is comparable in terms of age, marital status, and the type of house owned. Regional distributions also differ only modestly.\footnote{The comparison has been based on structural data on house owners in Germany available in the best for planning tool (b4p) (https://mds.mds-mediaplanung.de/). A complete overview of the comparison is outlined in the Supplementary Material B.} This testifies to the robustness of the dataset. Men are slightly over-represented within the sample (Figure \ref{fig:composition_gender}), and study respondents are more likely to have a high household income and to be highly educated than house owners in general. This is especially true for the sub-group CAs (Figure \ref{fig:composition_income}). Since PV systems are quite cost-intensive and can be assumed to appeal to house owners with some level of technical affinity, these findings are consistent with the researchers' expectations. A similar finding appears on the level of Sinus-Milieus\textsuperscript{\textregistered}. %The Liberal Intellectual, Social-Ecological, Adaptive Pragmatic, and Performer milieus are over-represented in the sample as opposed to the general population in Germany, while the Liberal Intellectual, Social-Ecological, and Adaptive Pragmatic milieus are also over-represented in the sample as opposed to the population of house owners. 
Although the regional distribution across Germany appears quite representative, CAs are more likely to live in urban areas than PAs (Figure \ref{fig:composition_residence}). An overview of previous adopters identified within the respondents' different social peer groups is presented in Figures \ref{fig:composition_adopt_fam}, \ref{fig:composition_adopt_friend}, and \ref{fig:composition_adopt_neigh}. While most respondents have one or more PV adopters among their friends and acquaintances, as well as among their neighbours, around half of the respondents reported having no PV adopters among their families and relatives. Furthermore, on average, male respondents identified slightly more PV adopters within their social peer groups than did female respondents. A significant difference is visible for respondents with high as opposed to low incomes throughout all peer groups. Finally, CAs have more PV adopters within their social circles than PAs in general; however, PAs with a high intention to adopt PV show comparable numbers.

\begin{sidewaysfigure*}[h]
\centering
\subfloat[Gender distribution]{\includegraphics[width=0.33\textwidth]{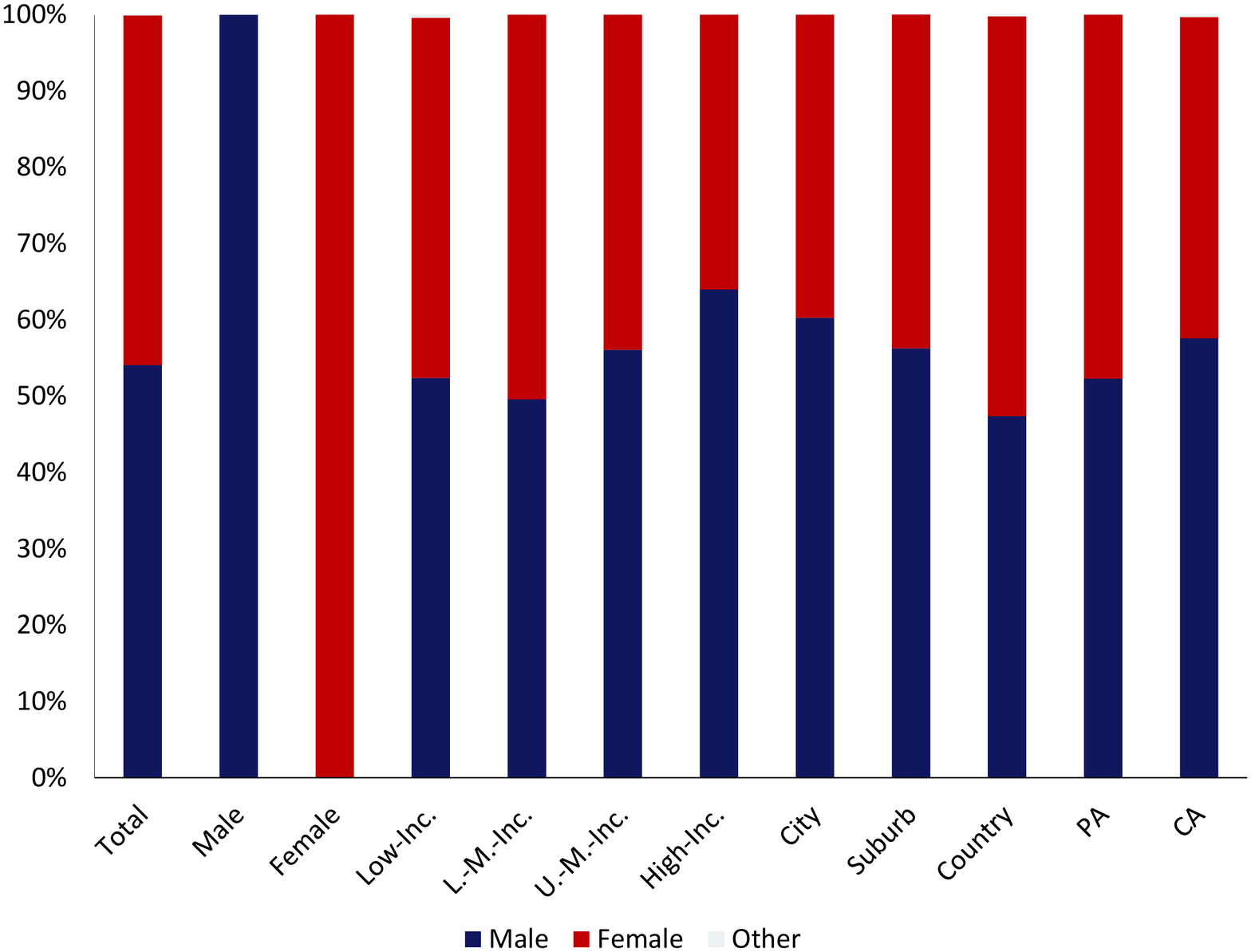}\label{fig:composition_gender}}\hspace{0cm}%  
\subfloat[Income level]{\includegraphics[width=0.33\textwidth]{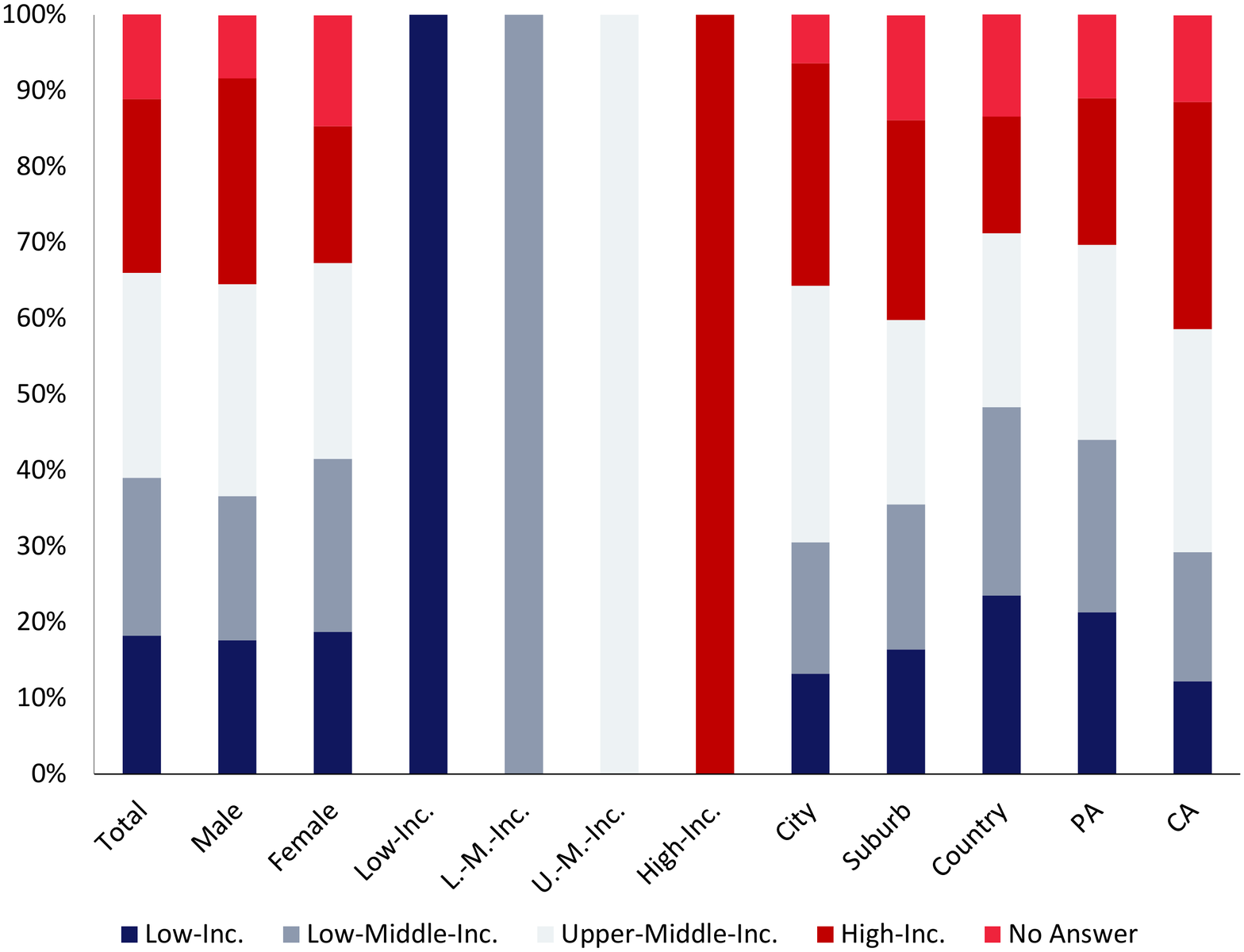}\label{fig:composition_income}}\hspace{0cm}%
\subfloat[Residence status]{\includegraphics[width=0.33\textwidth]{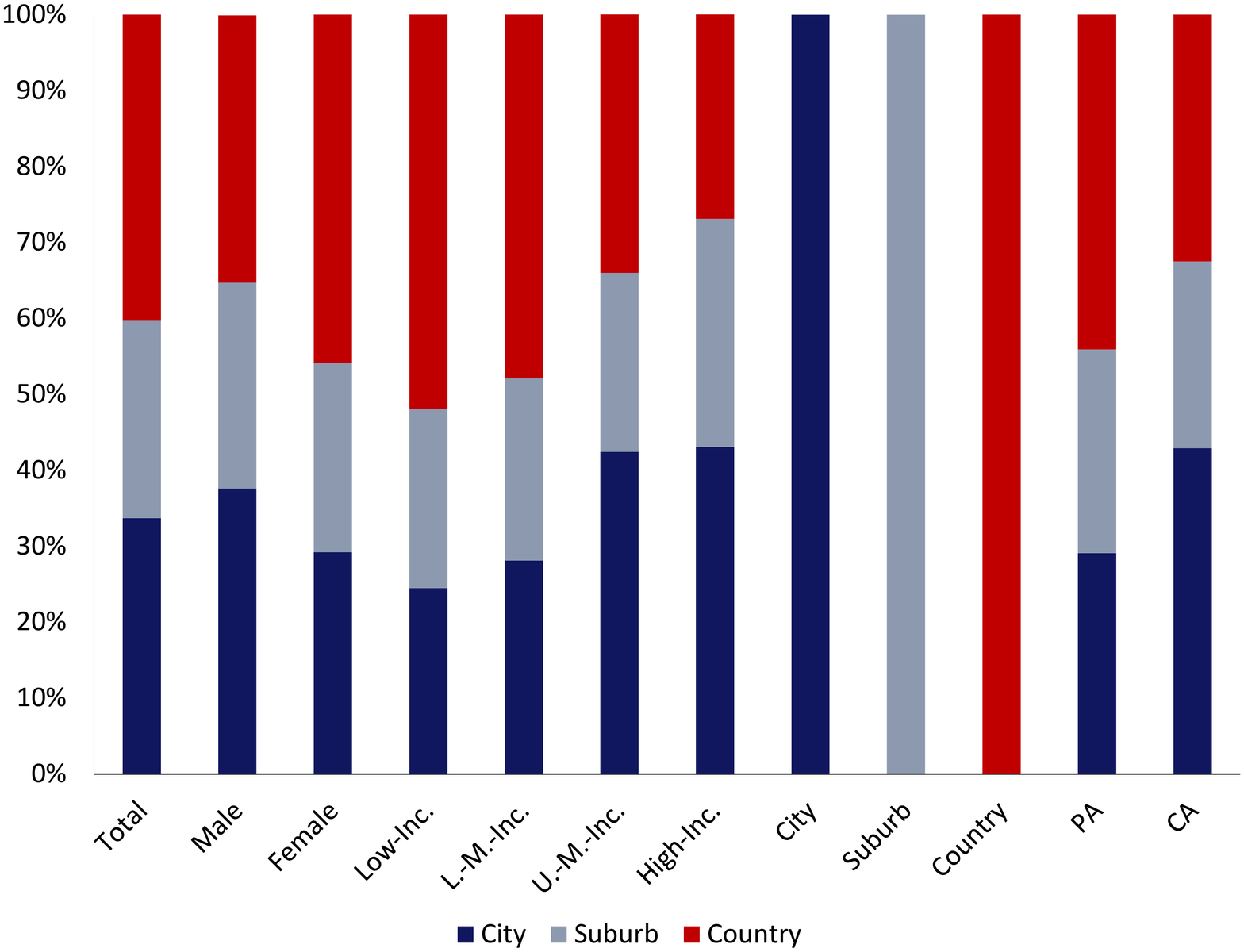}\label{fig:composition_residence}}\\ 
\subfloat[Adopters among family and relatives]{\includegraphics[width=0.33\textwidth]{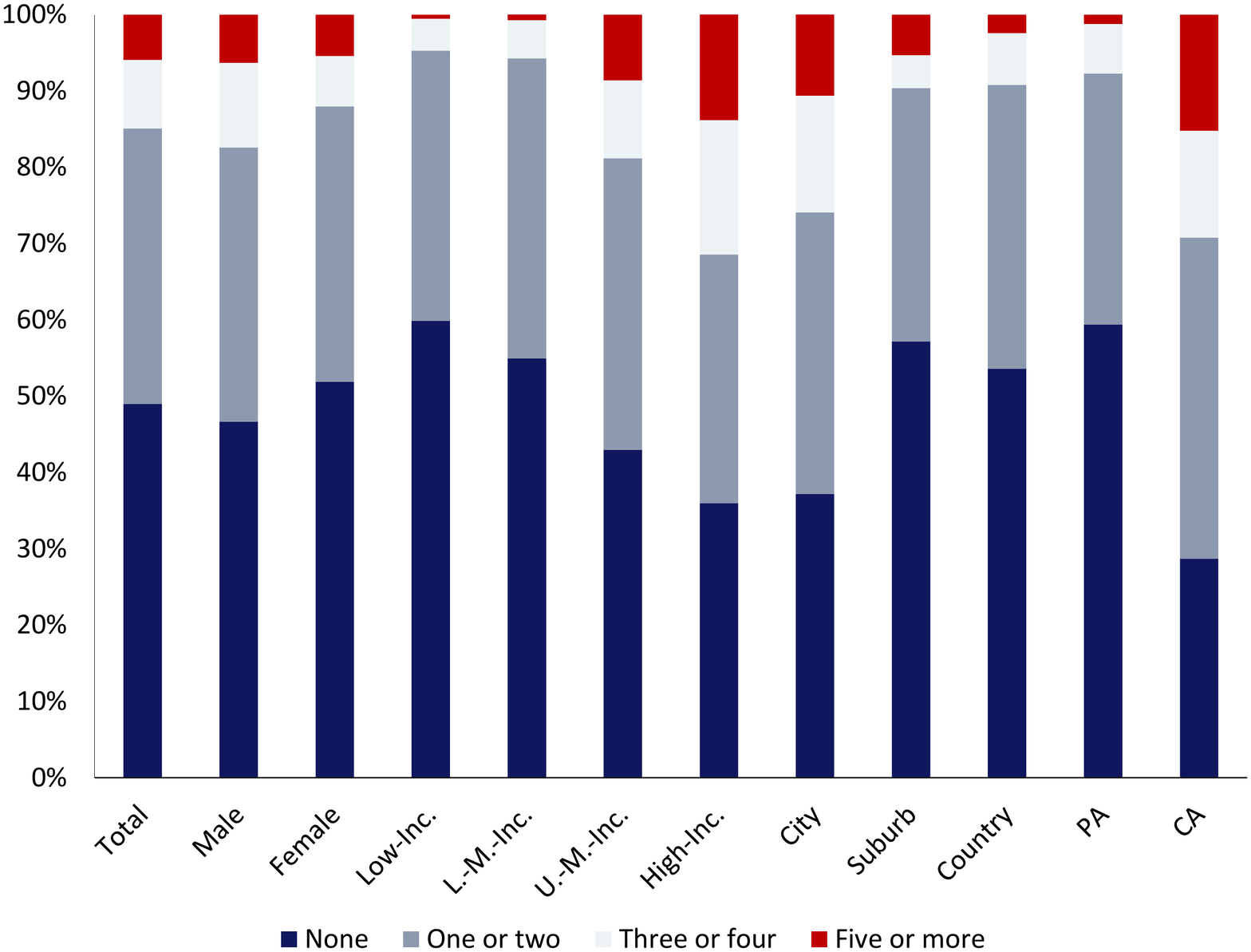}\label{fig:composition_adopt_fam}}\hspace{0cm}%  
\subfloat[Adopters among friends and acquaintances]{\includegraphics[width=0.33\textwidth]{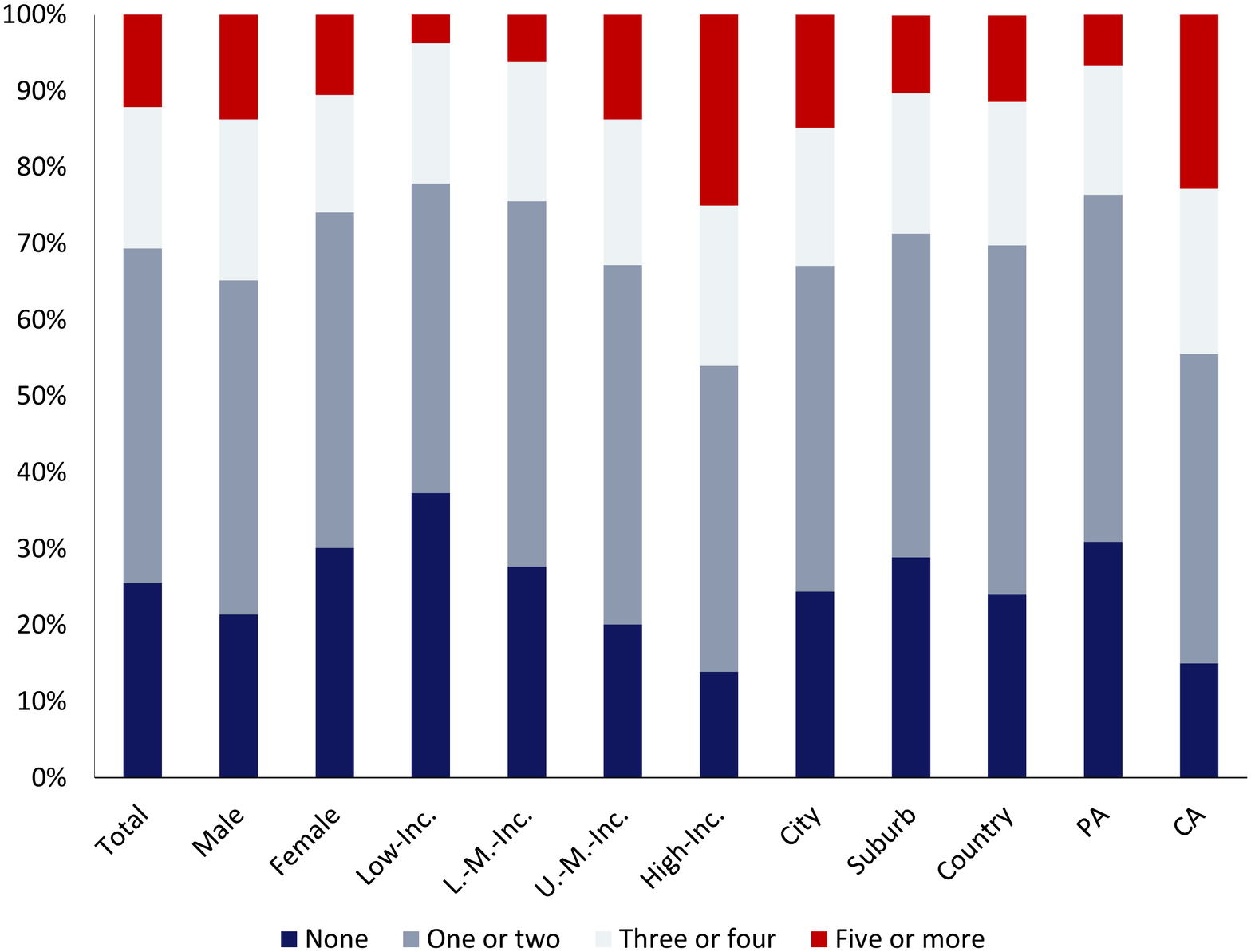}\label{fig:composition_adopt_friend}}\hspace{0cm}%
\subfloat[Adopters among neighbours]{\includegraphics[width=0.33\textwidth]{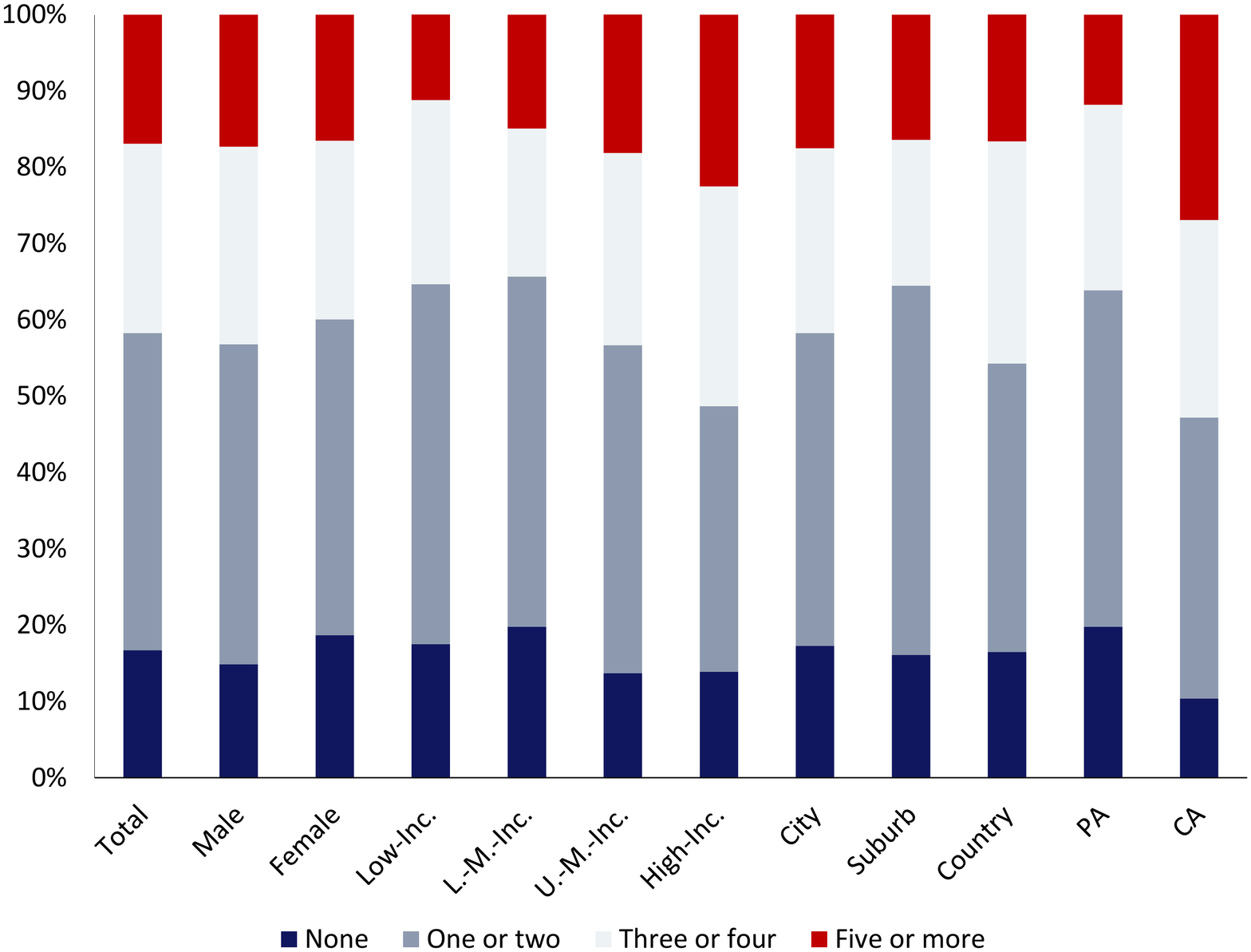}\label{fig:composition_adopt_neigh}}\\ 
\caption[SampleComposition]{Relative composition of the total sample (\textit{N=1,165}) in terms of personal characteristics of the decision-makers, such as gender, income level, and residence status (upper half). The total sample consists of more male respondents than female respondents; more upper-middle- to high-income respondents than lower-middle- to low-income respondents; and more respondents with a residence in the country than in the city. Furthermore, insights regarding the number of previous PV adopters in respondents' social peers groups are shown (lower half). While most of the respondents reported at least one adopter in their neighbourhood and at least one adopter among their friends and acquaintances, around half of the total sample indicated no adopters at all among their family members. Only slight differences are visible between genders and rural vs. urban residents; however, considerable differences are visible between income groups.} %Every figure includes the results for the values 10 \% - 30\% of the DR parameter share of total shiftable load. The DR parameter load shift horizon is varied between figures (a) to (d).}
\label{fig:sample_compositions}
\end{sidewaysfigure*}
\clearpage

\section{Investigation findings}
\label{S:4}

\subsection{Peer group pressure}
\label{sec:pressure}
The residential PV attitudes statement battery is presented in Section \ref{S:3.1}. The battery included eight statements about behavioural beliefs (economic (S1), autonomy (S2), environmental (S3), and control (S4) benefits of adopting PV), normative beliefs (social benefits through technology first-adopter status (S5), environmental consciousness (S6), and norm compliance (S7)), and social learning (satisfied adopters within peer groups (S8)) and is outlined in Table \ref{tab:statements}\footnote{The interrelation between PV statements recommended them as a candidate for factor analysis. Accordingly, principle component analysis was conducted. The most plausible finding was a Varimax-rotated four factor solution, which showed however, as cross-loadings are heavy, the model does not provide a significantly more parsimonious explanation than the original statement battery.}. 

\begin{table}[h!]
\footnotesize
\renewcommand{\arraystretch}{1.2}
\setlength{\tabcolsep}{5pt}
\caption{Descriptive statistics of respondents' agreement or disagreement with the following residential PV attitude statements. The means are related to a five-point ordinal scale (1=I disagree completely to 5=I agree completely). The sample (\textit{N=1,165}) consists both sub-groups PAs (\textit{n=394}) and CAs (\textit{n=771}), as presented in separate columns. While most of the respondents (\textit{n=1,065}) reported at least one or more adopters in their social peer groups (family and relatives, friends, acquaintances and co-workers, and neighbours) there are also some respondents (\textit{n=141}) who stated that they do not have any adopters in these groups. The descriptive statistics indicated a weaker average (mean) agreement with the PV attitude statements for respondents with no adopters in their social peer groups. The calculated standard deviations (sd) within this group are also higher on average, indicating more variation in responses.}
    \label{tab:statements}
\begin{tabularx}{\textwidth}{p{0.05\textwidth}| p{0.42\textwidth}| p{0.1\textwidth} p{0.1\textwidth}| p{0.1\textwidth} p{0.1\textwidth}}
\midrule
\textbf{Nr.} & \textbf{Residential PV-valuation statements} & \multicolumn{2}{p{3.7cm}|}{\textbf{PV adopters in peer groups}} & \multicolumn{2}{p{3.7cm}}{\textbf{No PV adopters in peer groups}}  \\
& &   mean & sd &  mean & sd\\
\midrule
S1 & A private PV system will save one money in the long term.  & 4.104 & .865 & 3.780 & .960\\
\hline
S2 & A private PV system will make one independent from energy providers. & 3.723 & 1.037  & 3.370 & .960\\
\hline
S3 & PV systems are good for the environment and the climate. & 
 4.210 & .817 & 3.970 & .989 \\
\hline
S4 & I can decide myself what technology gets installed on my roof. & 4.438 & .782 &  4.16 & 1.022 \\
\hline
S5 & Owning a PV system shows that one is interested in the latest technologies. &  3.705 & .987 &  3.36 & 1.059  \\
\hline
S6 & Owning a PV system shows that one is concerned about the environment and the climate. & 3.953 & .930 & 3.76 & 1.138  \\
\hline
S7 & My social circle would see owning a PV system as a good thing.  &  3.672 & .995 & 3.000 & .995\\
\hline
S8 & I know people who own PV systems and are satisfied with them. & 3.980 & 1.100 & 2.46 & 1.351 \\
\bottomrule
\end{tabularx}
\end{table}

Among both sub-groups PAs (\textit{n=771}) and CAs (\textit{n=394}), moderate to strong positive Spearman’s rank correlation coefficients were found between statements regarding the environmental, economic, autonomy, and social benefits of adopting PV. The strongest correlation was found between the environmental benefit statement and the environmental consciousness statement (\textit{\(\rho\)=.639***}\footnote{The significance levels are indicated as follows: \textsuperscript{*}$p<.050$, \textsuperscript{**}$p<.010$, \textsuperscript{***}$p<.001$.}). This suggested a good consistency within the response patterns in general. 
%Furthermore, a Mann-Whitney U test revealed that to a statistically significant degree, CAs are more likely than PAs to strongly or somewhat agree on all eight statements. 
Among PAs, both self-assessed location in the PV adoption decision process and likelihood of PV adoption moderately to strongly correlates with all benefit statements (Spearman’s rank correlation). %The relationships were examined using Spearman’s rank correlation coefficient. 
 
PAs were also presented with five statements about specific dimensions of PV acquisition intention (acquisition intention\footnote{S9: I intend to acquire a PV system within the next three years.} (S9), time investment\footnote{S10: I have spent quite a bit of time learning about PV systems.} (S10), money investment\footnote{S11: I have spent money on advising or information about PV systems.} (S11), investment capability\footnote{S12: I would gladly install a PV system if I had the money.} (S12), risk assessment\footnote{S13: The risks and costs that go along with PV systems are too high for me.} (S13)). A summary of the percentage distribution of responses among different demographic groups and additional relevant analysis results is given in Supplementary Material C.

The relationship between the number of PV adopters in the respondents’ social peer groups 
%(summation of PV adopters among the family and relatives, neighbours, friends and acquaintances) 
and their beliefs about PV systems was tested using Spearman’s rank correlation coefficient. In line with the descriptive results of Table \ref{tab:statements}, statistically significant positive correlations were found between the number of PV adopters and all statements regarding the benefits of adopting PV. As expected, strong positive correlations were found regarding the drivers related to social learning (\textit{\(\rho\)=.488***}) and general normative belief (\textit{\(\rho\)=.418***}). Slight to moderate positive correlations were found with more specific normative beliefs such as environmental concern (\textit{\(\rho\)=.218***}), and technological leadership (\textit{\(\rho\)=.278***}). Slight to moderate positive correlations were also found with statements regarding the environmental (\textit{\(\rho\)=.182***}), economic (\textit{\(\rho\)=.256***}), and autonomy benefits (\textit{\(\rho\)=.241***}) of PV adoption. Finally, a slight but statistically significant correlation was also found with the control belief (\textit{\(\rho\)=.162***}). Here, a plausible overall interpretation is that individuals who live in a PV-saturated social circle are both more susceptible to social pressure regarding PV adoption and more positively inclined toward PV in general. 

Among PAs, this is further supported by a moderate positive correlation between the number of PV adopters in the respondents’ social peer groups and the self-assessed likelihood of adopting PV (\textit{\(\rho\)=.252***}), as well as strength of intention to adopt PV within the next three years (\textit{\(\rho\)=.259***}). Plausibly, self-assessed likelihood of adopting PV and intention to adopt PV also correlate strongly with one another (\textit{\(\rho\)=.671***}). There is furthermore a moderate positive correlation between the number of PV peer adopters and the amount of time invested in the PV decision process (\textit{\(\rho\)=.262***}); a slight positive correlation with the amount of money invested in the PV decision process (\textit{\(\rho\)=.137**}); and a slight negative correlation with the assessment of risks and costs involved in adopting PV (\textit{\(\rho\)=-.160***}). 

The relationship between the area of residence (city, suburb, country) and the PV statements among CAs and PAs was examined using the Kruskal–Wallis H test. Statistically significant relationships were found concerning all statements except the satisfaction and control statement. The Mann–Whitney U test was then utilized post-hoc to compare each area of residence. 
%In this context, urban respondents are the most likely to agree with several statements regarding the normative benefits of PV systems. 
Urban respondents were especially more likely than suburban and rural respondents to agree with the statements regarding technological leadership, acceptance within social peer groups, and environmental concern.
%Thereby, at a statistically significant level, urban respondents are more likely than suburban and rural respondents to agree with the statements regarding technological leadership (S5: p$<$.001) and acceptance within social peer groups (S8: p$<$.001). 
%Additionally, at a statistically significant level, urban respondents are more likely than rural respondents to agree with statements regarding the environmental concern (S6: p$<$.001). The like applies to the reporting of economic (S3: p=.012), environmental (S1: p=.012), and autonomy (S2: p$<$.001) benefits. 
No statistically significant differences in patterns of agreement with the statements were found between suburban and rural respondents. %In brief, urban respondents are the most likely to agree with several statements regarding the normative benefits of PV systems. 
It cannot be argued, however, that urban respondents are more likely to know social peers who own PV systems and are satisfied therewith, nor that they feel as if they have greater control over domestic PV adoption. The same goes for perceived neighbourhood cohesion in correlation to the statements which were tested using Spearman’s rank correlation coefficient. While better perceived neighbourhood cohesion leads to slight, yet statistically significant positive correlations with respect to all benefit statements, the correlations are simply too slight to support clear conclusions.

%(e.g.,S1: \textit{\(\rho\)=.152***; S2: \textit{\(\rho\)=.135***})S3: \textit{\(\rho\)=.149***; S4: \textit{\(\rho\)=.079**}), the normative belief statements (S5: \textit{\(\rho\)=.128***; S6: \textit{\(\rho\)=.159***; S7: \textit{\(\rho\)=.180***}), and the social peer acceptability statement (S8: \textit{\(\rho\)=.128) 

Despite the high mean and median income within the sample as a whole, a Spearman’s rank correlation test demonstrated a statistically significant positive correlation between income and the number of PV adopters in the respondents’ social peer groups (\textit{\(\rho\)=.256***}). Slight but statistically significant correlations were also found between income and perceived economic (\textit{\(\rho\)=.110***}), autonomy (\textit{\(\rho\)=.143***}), and control (\textit{\(\rho\)=.105***}) benefits; normative beliefs (\textit{\(\rho\)=.146***}); norm compliance (\textit{\(\rho\)=.171***}); and social acceptability (\textit{\(\rho\)=.105***}). Within the sample of individuals with an interest in PV, it is reasonable to assume that those with the financial means to adopt a PV system may harbour more positive attitudes toward PV, as they do not face a financial barrier. However, it is difficult to disambiguate whether the observed differences are due to income, or rather due to the collinear relationship between income and the number of peer adopters in the social circle.

With all of these results in mind, two linear regression analyses among PAs were applied. The aim was to assess the capacity of the PV attitude statements to predict, first, self-reported location in the PV decision-making process, and second, self-reported likelihood of adopting PV. The tests were carried out as a step-wise regression to single out non-significant drivers and barriers (\textit{p$<$.005}) and was controlled for the number of adopters in the peer groups, neighbourhood cohesion, income level, and gender. Regarding self-assessed location in the PV decision-making process (\textit{$R^2$=.188, n=686}), the $\beta-$coefficients demonstrated a small but significant influence of the number of PV peer adopters (\textit{$\beta_{adopter}$=.023**}) as well as positive influences of the statements on time investment (\textit{$\beta_{S10}$=.097***}), money investment (\textit{$\beta_{S11}$=.080***}), investment capability (\textit{$\beta_{S12}$=.346*}), and environmental concern (\textit{$\beta_{S6}$=.356*}). Regarding self-assessed intention to adopt (\textit{$R^2$=.382, n=686}),  the number of peer adopters played a positive but non-significant role; amount of time and money invested, investment capability, risk assessment were explanatory variables (\textit{$\beta_{S10}$=.239***; $\beta_{S11}$=.182***; $\beta_{S12}$=.081**; $\beta_{S13}$=-.117***}), as was income level (\textit{$\beta_{Inc}$=.041**}). The model is completed by benefit statements about long-term economic benefit (\textit{$\beta_{S1}$=.183***}), independence (\textit{$\beta_{S2}$=.111**}), satisfaction (\textit{$\beta_{S8}$=.055*}), and acceptance (\textit{$\beta_{S7}$=.126**}). Taken together, the regression analyses suggest that some caution should be taken when assessing the effect of peer influence on the likelihood of adopting PV, since variables such as income level, investment capability, and perceived long-term economic benefit also play an important explanatory role.%.% on the motivation of respondents to adopt or intention to adopt PV. Despite the high-income level in general across the sample in this survey, variables such as the income level, investment capability, perception of long-term economic benefit played an important role. %for the explanation. % of the intention and should therefore not be left out of the evaluation.    

\subsection{Peer group perception}
\label{sec:perception}

\begin{sidewaysfigure*}
\centering
\subfloat[Trustworthiness]{\includegraphics[width=0.25\textwidth]{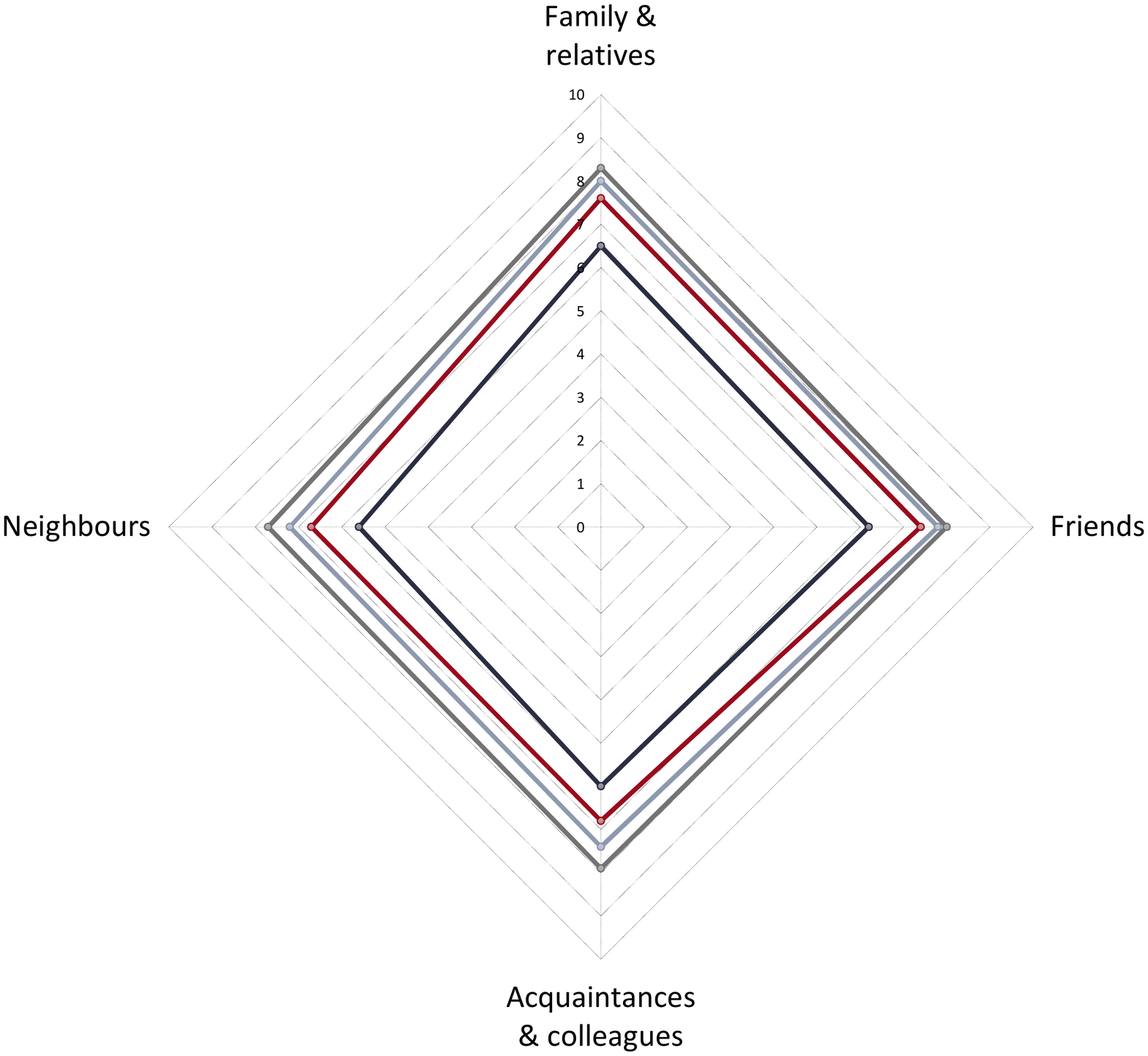}}\hspace{0cm}%  
\subfloat[Competence]{\includegraphics[width=0.25\textwidth]{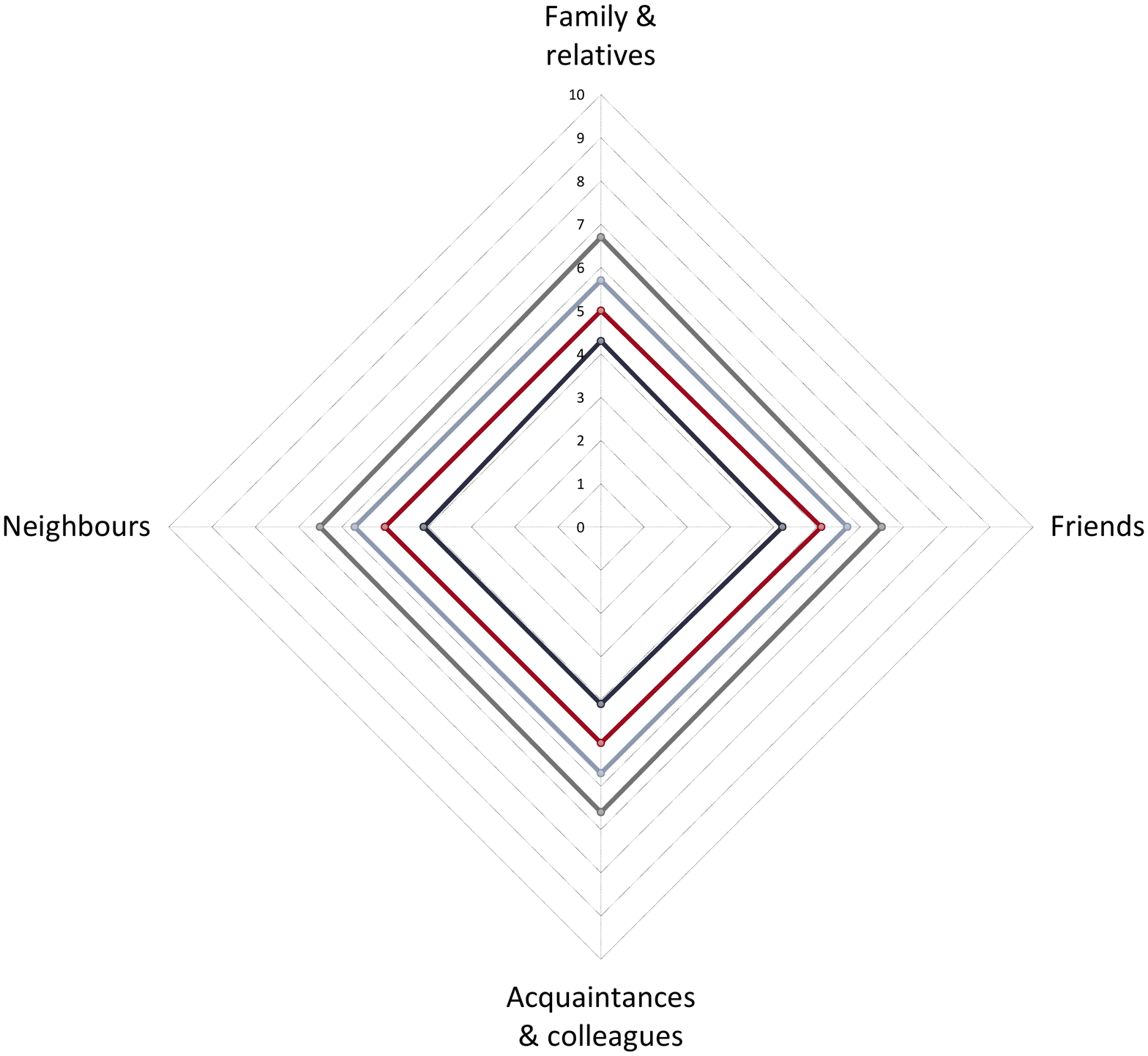}}\hspace{0cm}%
\subfloat[Power]{\includegraphics[width=0.25\textwidth]{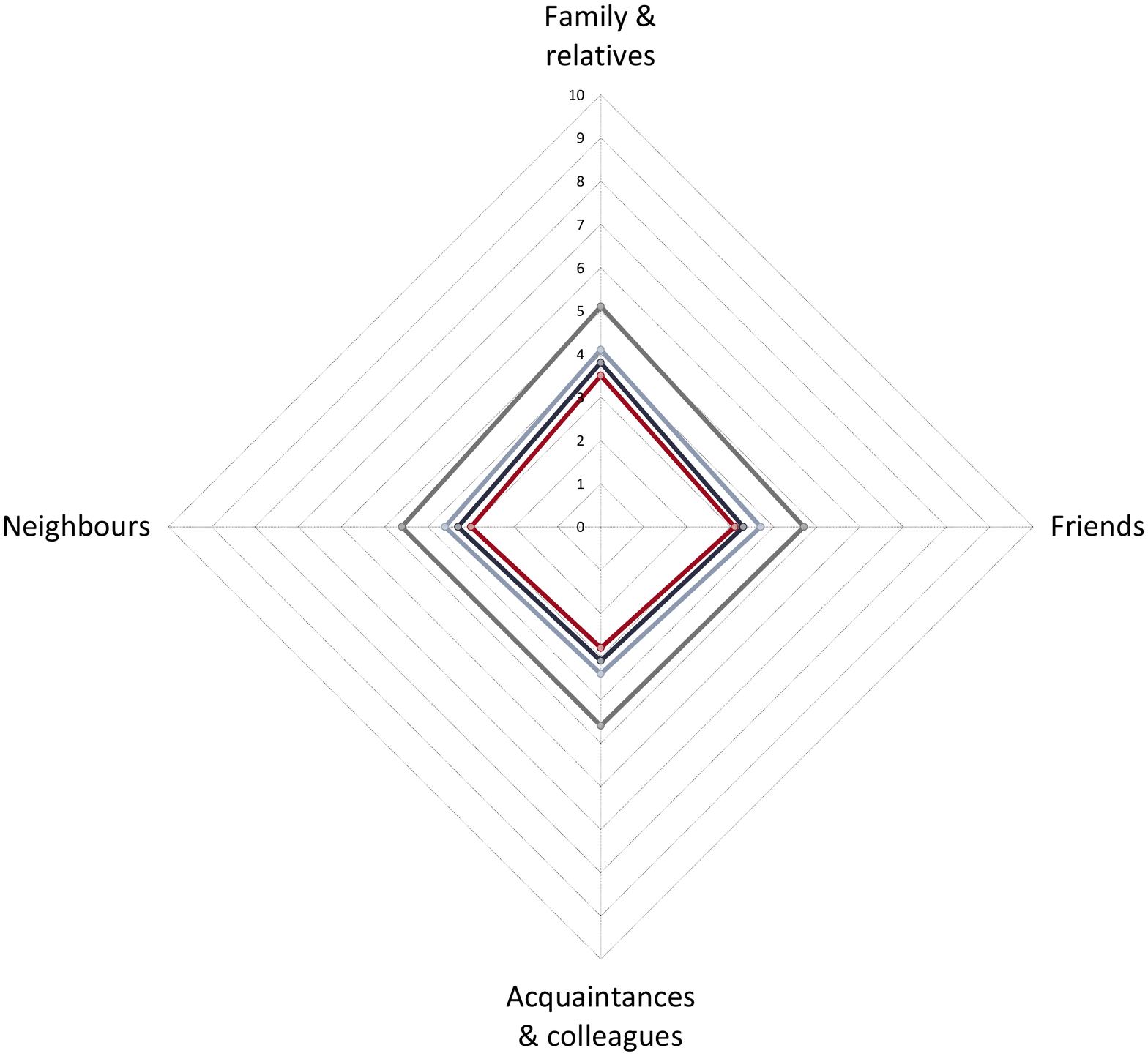}}\\ 
\subfloat[Independence]{\includegraphics[width=0.25\textwidth]{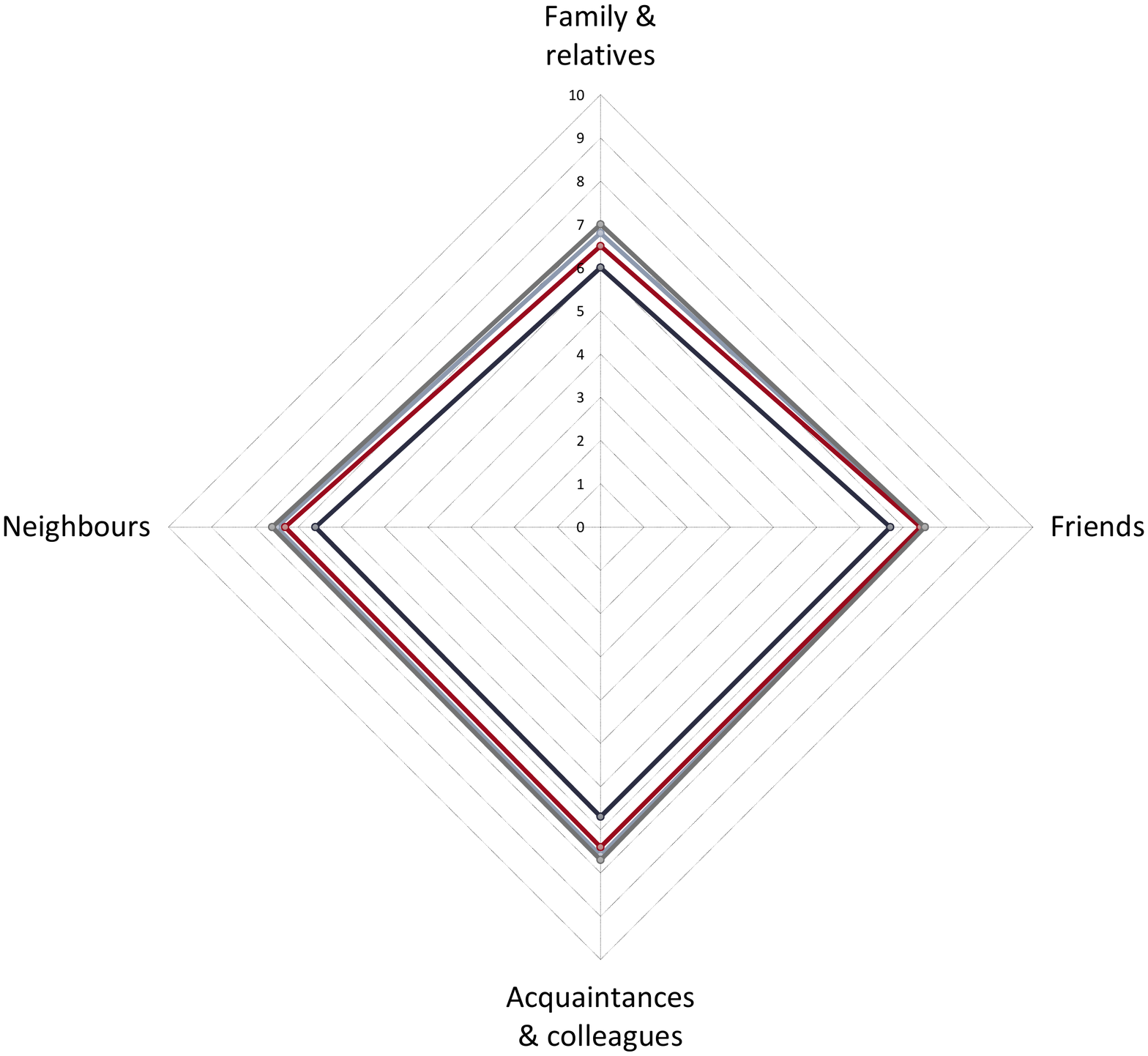}}\hspace{0cm}%  
\subfloat[Availability]{\includegraphics[width=0.25\textwidth]{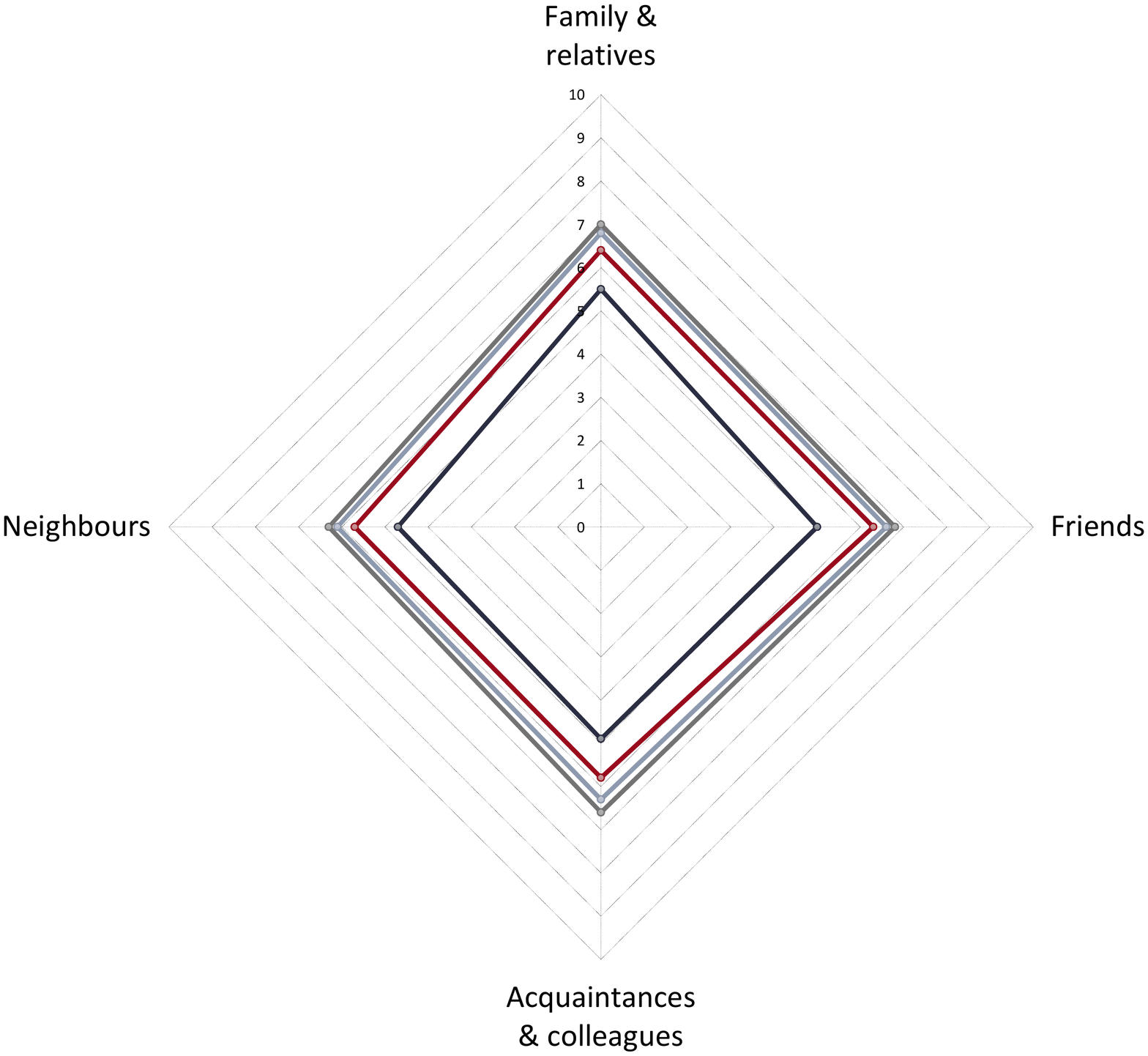}}\hspace{0cm}%
\subfloat[Closeness]{\includegraphics[width=0.25\textwidth]{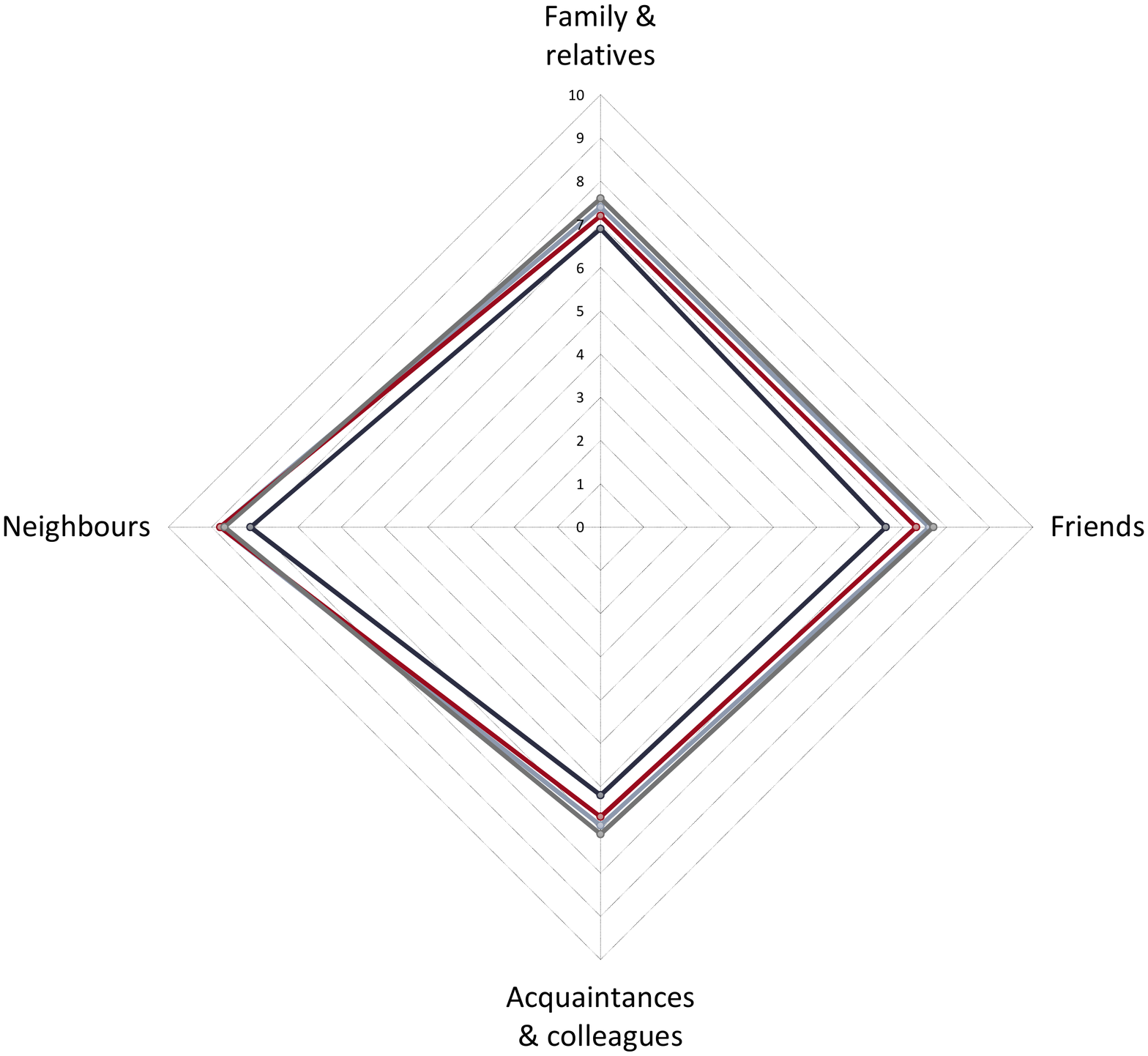}}\\ 
\subfloat[likeability]{\includegraphics[width=0.25\textwidth]{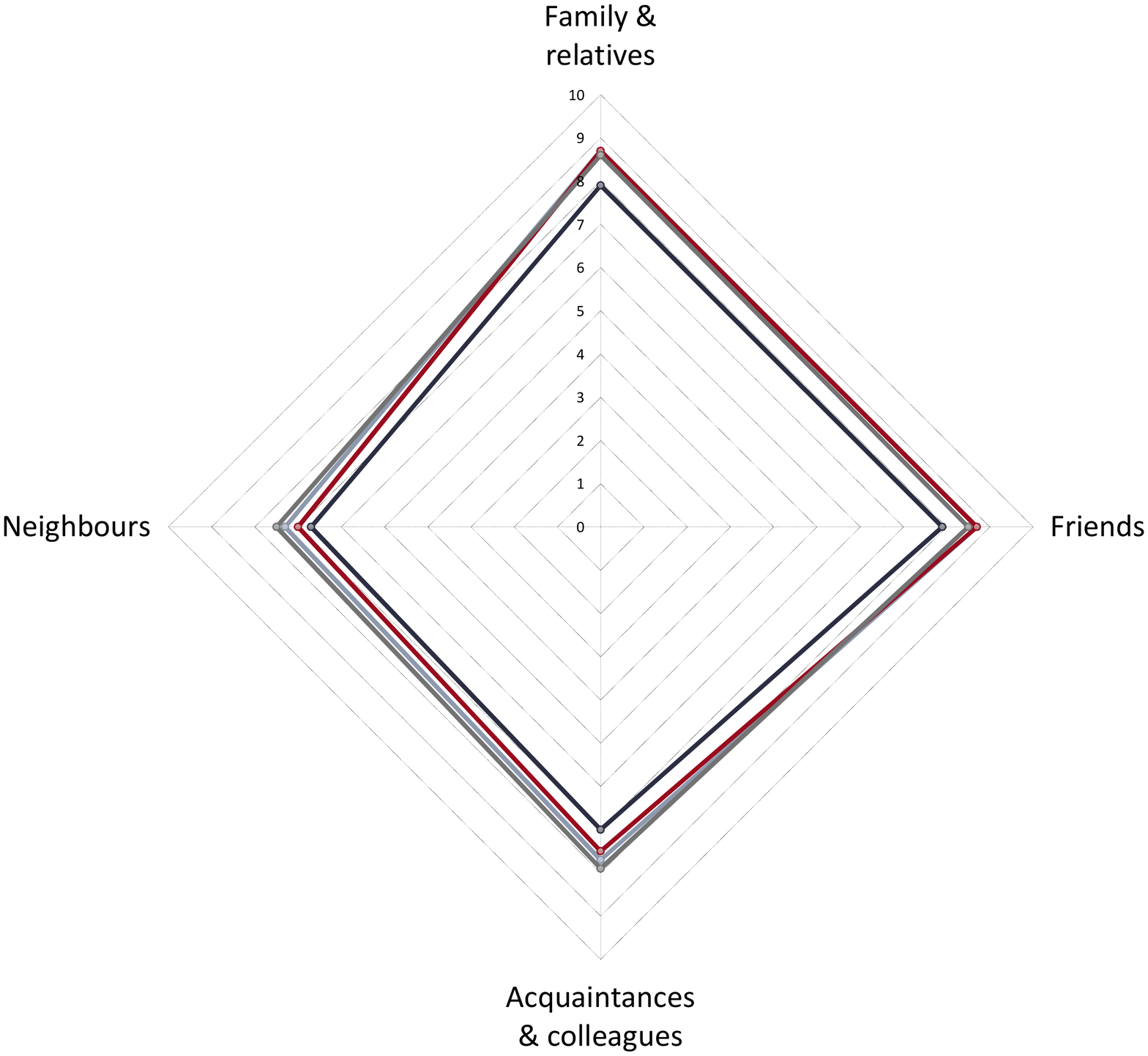}}\hspace{0cm}%
\subfloat[Integrity]{\includegraphics[width=0.25\textwidth]{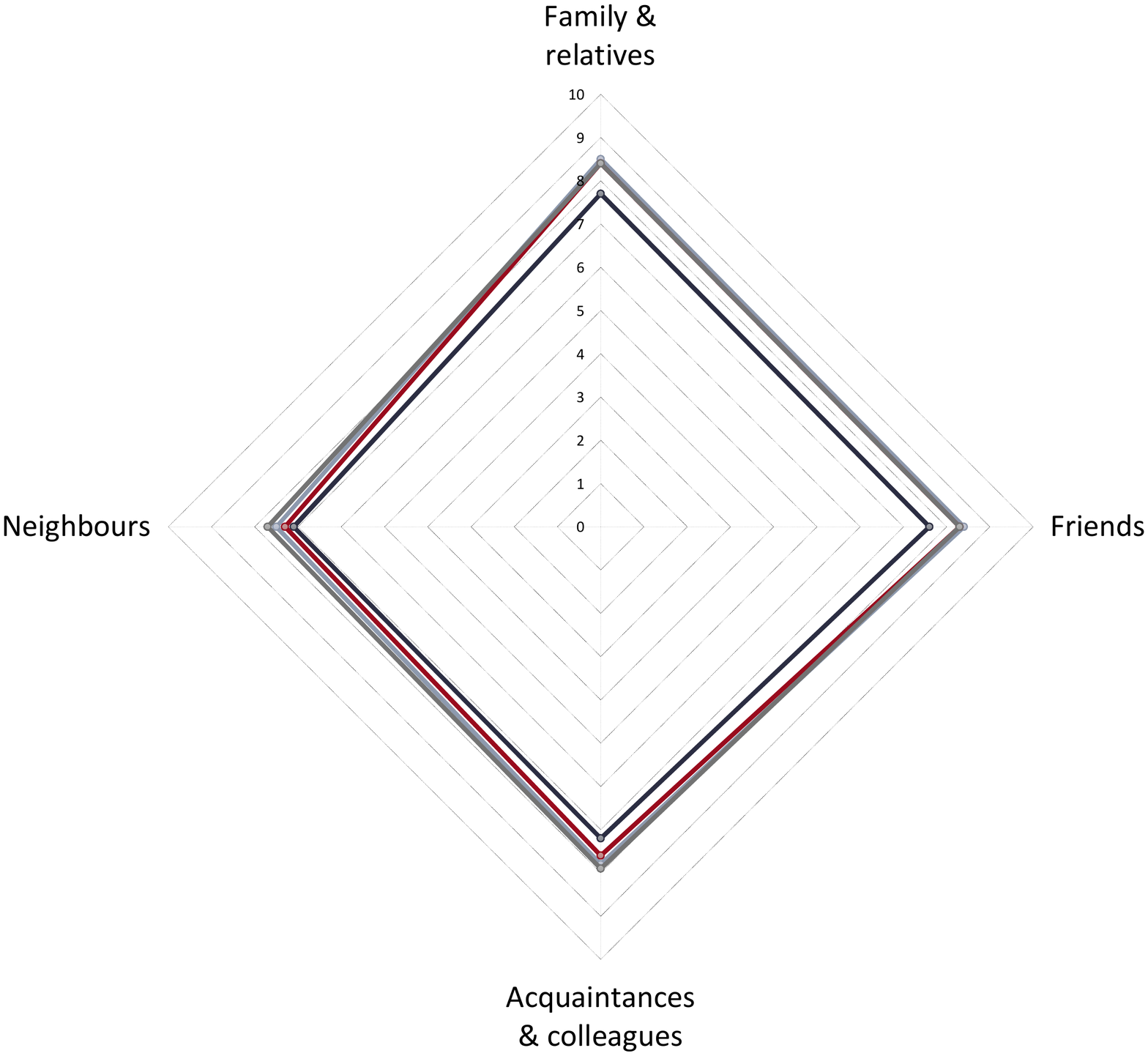}}\hspace{0cm}% 
\subfloat[Reliability]{\includegraphics[width=0.25\textwidth]{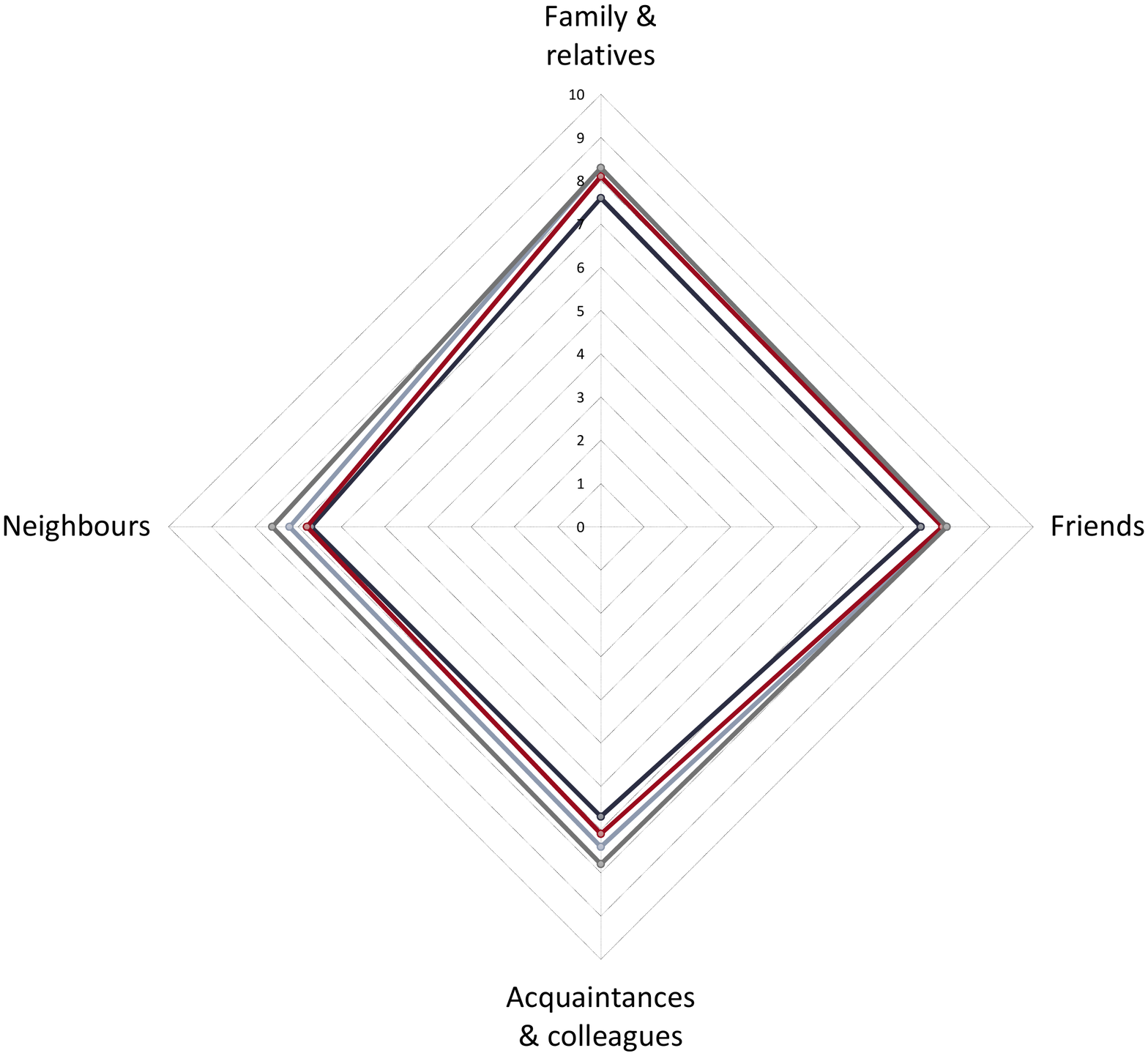}}\\
\hfill\subfloat{\includegraphics[width=0.4\textwidth]{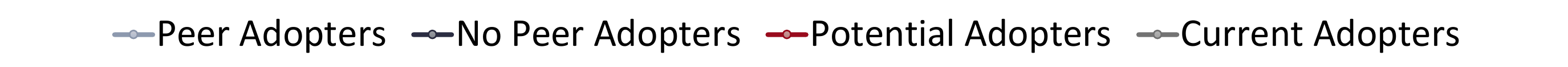}}
\caption[Perception]{Descriptive statistics of the perceptions of attributes of social peer groups with respect to the PV adoption process. The presented means represent scores on a sliding scale (1=the selected attribute does not apply at all to the respective peer group to 10=the selected attribute applies completely to the respective peer group). The samples consist of the sub-groups PAs (\textit{n=394}) and CAs (\textit{n=771}), as well as the sub-groups of respondents (n=1,065) with one or more adopters in their social peer groups (family and relatives, friends, acquaintances and acquaintances, as well as neighbours) and respondents without any adopter in their social peer groups (n=141). The descriptive statistics indicated a weaker average rating of attributes for peer groups with no adopters in their social peer groups. The attributes trustworthiness and competence showed the greatest differences. The highest total ratings among the perceived attributes were given to neighbours for closeness, family members and friends for likeability, and acquaintances for integrity.}
\label{fig:peer_perception}
\end{sidewaysfigure*}

Decision-makers are confronted, at different stages in the PV decision-making process, with various social peers exhibiting different attributes. Since individual decision-makers might perceive peers in different ways, survey respondents were asked to rate peers within four groups along nine attributes as defined in Table \ref{tab:Terms}\footnote{Similarly to the statement battery, the interrelation of attributes suggested that latent factors may be identifiable. Accordingly, a principal component analysis of attributes across all actors was conducted. Various factor solutions were tested; a Varimax-rotated five-factor solution appeared plausible, explaining 87.37\% of variance in actor attribute ratings through the following coherent groupings of attributes. After various tests, it was decided to leave the original variables for the following reasons:
1) the attribute likeability did not load cleanly to any single factor while representing a crucial variable in existing literature but also in this analysis regarding the underlying mechanisms of active peer effects;  
2) three sets of dimensions reduced well to single factors: trustworthiness and competence, availability and distance, and integrity and reliability; however, it was decided that maintaining the ability to disambiguate trustworthiness and competence in particular was important from a theory and modeling standpoint;
3) finally, it was not significantly more parsimonious to utilise five constructs as opposed to nine; it was decided that a slight gain in parsimony does not balance out the losses in granularity explained above.}. 
While an initial descriptive indication is outlined in Figure \ref{fig:peer_perception} for respondents who have PV adopters in their social peer groups and for respondents who do not, as well as for the sub-groups PAs and CAs, the results of the Spearman’s rank correlation coefficient analysis are described in the following section.

The highest average rating over the entire sample was given to neighbours for the attribute closeness (\textit{$8.744\pm1.839$, n=1,122}), as well as family members and friends for the attribute likeability (\textit{$8.656\pm1.856$, n=1,127; $8.609\pm1.794$, n=1,126}). The lowest ratings were given on the attribute power for all of the peer groups. Considerable differences in mean ratings for the attributes trustworthiness and competence were found between respondents with and respondents without PV adopters in their social peer groups. 

The survey revealed that the number of PV adopters in a respondents’ social peer groups had a clear impact on the reported perceived attributes of the respective social peer group. For instance, statistically significant positive correlations were found between the number of PV adopters among the respondents’ neighbours and respondents’ ratings of neighbours’ PV-relevant attributes. The strongest correlations were found regarding the attributes trustworthiness (\textit{\(\rho\)=.328***}), competence (\textit{\(\rho\)=.308***}), and availability (\textit{\(\rho\)=.295***}). At the same time, the attributes power, independence, and closeness showed negligible correlations. This seems plausible, as these attribute ratings should not depend upon whether or not neighbours have adopted PV systems. 
The relationship between the number of PV adopters within the respondents’ families and respondents’ ratings of family members’ PV-relevant attributes showed a similar result. As expected, statistically significant positive correlations were found with most attribute ratings. The strongest correlations appeared with the attributes competence (\(\rho\)=.526***), availability (\(\rho\)=.312***) and power (\(\rho\)=.284***). As expected, the following attributes did not show statistically significant correlations: independence, likeability, integrity, and reliability. %This seems plausible, as these attribute ratings should not depend upon whether or not family members have adopted PV systems.

%Independence (p$=$.482), likeability (p$=$.596), integrity (p$=$.643), reliability (p$=$.421). This is plausible, as these attribute ratings should not depend upon whether or not family members have adopted PV systems.

Finally, with regard to respondents' friends group and co-worker/acquaintances group, 
%slight to moderate, statistically significant correlations were found with most attribute ratings. 
the number of PV adopters was found to correlate most strongly with the attribute competence (friends: \(\rho\)=.367***; acquaintances: \(\rho\)=.382***), followed by trustworthiness (friends: \(\rho\)=.322***; acquaintances: \(\rho\)=.352***). 
The attribute independence for both friends and acquaintances and the attribute likeability for friends were not found to correlate with the number of PV adopters at a significant level. This seems plausible, as whether or not one’s friends or acquaintances own a PV system should not impact one’s assessment of their independence or likeability. Further results of the analyses are shown in Supplementary Material D.

%Exceptions are the attribute independence for both friends (\(\rho\)=-.007, p$=$.823) and acquaintances (\(\rho\)=.001, p$=$.971) and the attribute likeability for friends (\(\rho\)=.056, p$=$.058), neither of which correlate with the number of PV adopters within the friend and co-worker group at a significant level. This is plausible, as to whether or not one’s friends or acquaintances own a PV system should not impact one’s assessment of their independence or a friend’s likeability.

\subsection{Peer group contact initiation}
\label{sec:communication}

\begin{figure*}[h]
\centering
\subfloat[Reported contacts in decision-making stage I]{\includegraphics[width=1\textwidth]{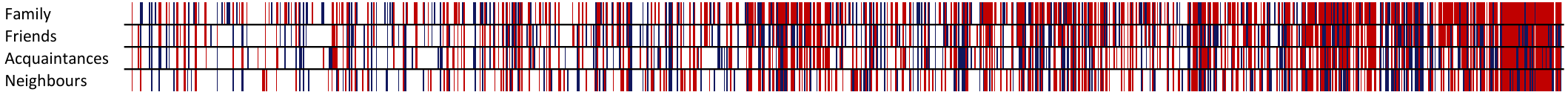}}\hspace{0cm}%  
\\ 
\subfloat[Reported contacts in decision-making stage II]{\includegraphics[width=1\textwidth]{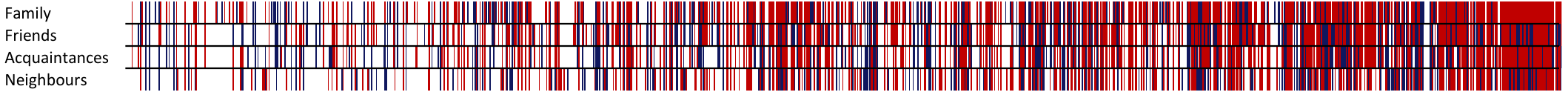}}\hspace{0cm}%    
\\ 
\subfloat[Reported contacts in decision-making stage III]{\includegraphics[width=1\textwidth]{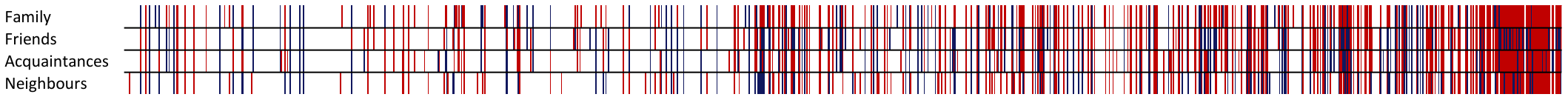}}\hspace{0cm}%    
\\ 
\hfill\subfloat{\includegraphics[width=1\textwidth]{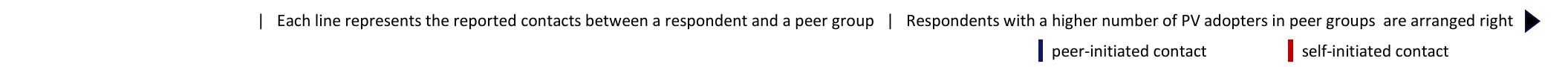}}

\caption[Contact]{Illustrative overview for the reported contacts of the respondents with varying social peer groups. Each line represents the answer of one respondent in one of the stages of decision-making. The respondents are sorted in ascending order according to the number of adopters in their peer group (respondents on the left stated hardly any PV adopters in their peer groups; respondents on the right stated various PV adopters in their peer groups). While an actively initiated contact between the decision-maker and one of the social peer groups is highlighted in red, a passively experienced contact or a contact where the respondent is not sure anymore is highlighted in blue. Thus, the white lines are depicting no contact or no answer from the the respondents at all. The samples for the different stages consist of the sub-groups PAs (\textit{n=394}) and CAs (\textit{n=771}). Since only CAs and those PAs who reached decision-making stage III had been requested to report the contacts for the last decision-making stage, the observations were lower for stage III (n=563) than for stage II (\textit{N=1,165}) and stage I (\textit{N=1,165}). The figure indicates a higher rate of actively initiated contacts between respondents and social peer groups in case there is a higher number of PV adopters in their peer group. Furthermore, these respondents also experienced more contacts overall.}
\label{fig:exchangeinitiation}
\end{figure*}

Within each relevant decision-making stage, respondents were asked whether they came into contact with each social peer group, whether the information exchange was unidirectional or bidirectional, and who initiated the contact (Section \ref{S:3.1}). An overview of respondents' reported self-initiated and peer-initiated contacts in relation to the number of PV adopters in their social peer groups is given in Figure \ref{fig:exchangeinitiation}. Over the total sample (\textit{N=1,165}), around half of the respondents had contact with each of the peer groups in stage I (family: 56.4\%; friends: 51.2\%; acquaintances: 45.2\%; neighbours: 42.1\%). However, the rates of contact are lower for respondents (n=141) who did not report any any PV adopters in their peer group (family: 43.3\%; friends: 38.3\%; acquaintances: 34.8\%; neighbours: 29.8\%). The same trend is visible in stages II and III, as depicted in Figure \ref{fig:exchangeinitiation}. Moreover, the number of reported self-initiated contacts increased with the number of peer PV adopters throughout all three stages.

The relationship between the number of PV adopters among the respondents’ social peer groups and respondents’ likelihood of having communications of any type with these peers was checked using a count of frequencies and chi-squared test. During all three decision-making stages, at a statistically significant level, respondents with PV adopters among their neighbours were both more likely than average to have communicated in any form with their neighbours and more likely to have engaged in direct bilateral exchanges about PV with their neighbours. %(stage I: \textit{p$<$.001}; stage II: p$<$.001; stage III: p$=$.014). 
The same was the case with family, friends, and acquaintances, within all three decision-making stages and with regard to both communications in general and direct bilateral exchanges. These results are further supported by the fact that respondents with PV adopters in a particular peer group were significantly more likely than average to identify contact with that group as a spark event.

Since relationships were found between the number of PV adopters in a respondent's social peer groups and their likelihood of having adopted PV or intention to adopt, as well as between the number of PV adopters in a respondent's social peer groups and the perceived positive attributes of each group with respect to PV, it was hypothesised that the decision to initiate PV-related interactions might also depend on the perceived attributes of the available interlocutors. To examine which attributes of the social peer groups (Section \ref{sec:perception}) triggered the decision maker to actively seek contact to various peers viewed as a whole group, a logistic regression analysis was conducted. The results are displayed in the left panel of Table \ref{tab:AttributeImportance}. The dependent variable ’active initiation of peer exchanges’ comprises all exchanges  between the decision-maker and all of the peer groups which were initiated by the decision-maker herself or himself. A step-wise regression was applied to single out non-significant attributes (\textit{p$<$.005}). The regression was controlled for the number of adopters in the peer group as a whole and the respondent's area of residence. It was also controlled for the respondents' gender and  income level, neither of which was found to be significant. Note that the stated $\beta-$coefficients constitute the linear marginal effect of a unit increase of the independent variables on the odds ratio of the exchange initiation for the average respondent. With this remark in mind, the following observations can be made: having adopters in the peer group had a positive, high, and increasing impact over the stages (\textit{$\beta_{I}$=.131***; $\beta_{II}$=.149***; $\beta_{III}$=.252***}). In the first stage of the decision process, it was complemented by the effect of competence, which was of high impact and significance over all phases (\textit{$\beta_{I}$=.178***; $\beta_{II}$=.122***; $\beta_{III}$=.163***}). While availability, power and closeness had a minor impact, independence had a slight negative impact. Likeability was a particularly important driver in the first two stages (\textit{$\beta_{I}$=.138***; $\beta_{II}$=.201***}), whereas trustworthiness (\textit{$\beta_{III}$=.130***}) and reliability (\textit{$\beta_{III}$=.085**}) increased in relative importance in the third stage. This finding is plausible when taking into account that as a decision-maker advances through the stages, the decision process becomes more serious, and rational assessments gain importance relative to emotional inclinations. This interpretation is backed by the increase of the explained variation through the stages (\textit{$R^2$=.152; $R^2$=.153;$R^2$=.210}), when regarded as the non-random part of the communication decision. Lastly, living far from urban areas had a negative effect on the odds of actively seeking communication, which might be related to the density of PV adopters, as well as to urban respondents' generally more positive perceptions of PV.

\begin{table}[h!]\centering
\scriptsize
\caption{Impact of the indicated peer group attribute variables (trustworthiness, competence, power, independence, availability, closeness, likeability, integrity, reliability) and of the reported control variables (gender, income, residence, adopters) on the reported active initiation of exchanges (logistic regression) with the different groups of social peers (family and relatives, friends, acquaintances and co-workers, neighbours), as well as on the indicated influence strength (linear regression) of social peers in residential PV adoption decisions. On the basis of stepwise regressions, the impact of the explanatory variables on the dependent variable is assessed for the transformed responses matrix for each of the decision-making stages. Multicollinearity of the explanatory variables of each model has been checked and the table presents only coefficients with a $p<.05$ significance level. While the dependent variable with the header 'active initiation of peer exchanges' comprises all exchanges between peers and the decision-maker which have been initiated by the decision-maker herself or himself (logistic regression), the column with the header 'influence of self-initiated peer exchanges' comprises the influence exerted by the peers on the decision-maker as a result of these self-initiated exchanges (linear regression). Additionally, the column with the header 'influence of all peer exchanges' comprises the influence exerted by the peers on the decision-maker as a result of any exchange, whether self-initiated by the decision-maker or initiated by the peer (linear regression). In this paper, exchanges include both unidirectional reception of information and bidirectional exchanges, e.g. face-to-face conversations or email exchanges: an example of peer-initiated unidirectional contact is being fowarded a link to information about PV; an example of self-initiated unidirectional contact is seeking out and reading a blog entry about PV; an example of peer-initiated bidirectional contact is being approached by a peer for a conversation about PV; an example of self-initiated bidirectional contact is approaching a peer for a conversation about PV}
\begin{tabular}{l | l l l | l l l | l l l }
 & \multicolumn{3}{l |} {Logistic regression (Section \ref{sec:communication})} & \multicolumn{3} {  l | }{Linear regression (Section \ref{sec:influence})}& \multicolumn{3} {  l }{Linear regression (Section \ref{sec:influence})} \\ 
\toprule
\textbf{Explanatory}& \multicolumn{3} { l | }  {\textbf{Active initiations of}} & \multicolumn{3} {  l | }{\textbf{Influences of active initiated}}& \multicolumn{3} {  l }{\textbf{Influences of all experienced}} \\ 
\textbf{variables}& \multicolumn{3} { l | }  {\textbf{peer exchanges}} & \multicolumn{3} {  l | }{\textbf{peer exchanges}}& \multicolumn{3} {  l }{\textbf{peer exchanges}} \\ 
 & stage I & stage II & stage III & stage I & stage II & stage III & stage I & stage II & stage III \\ 
\midrule
Trustworthiness    &           &           & .130***   & .204***  &         & .214*** & .168***  & .124*** & .173*** \\
                   &           &           & (.039)    & (.042)   &         & (.056)  & (.034)   & (.033)  & (.042)  \\
Competence         & .178***   & .122***   & .163***   & .320***  & .347*** & .257*** & .247***  & .289*** & .220*** \\
                   & (.024)    & (.023)    & (.028)    & (.028)   & (.024)  & (.042)  & (.023)   & (.023)  & (.035)  \\
Power              & .076***   & .045**    & .088***   & .140***  & .110*** & .094*** & .132***  & .114*** & .117*** \\
                   & (.015)    & (.015)    & (.017)    & (.017)   & (.016)  & (.019)  & (.014{]} & (.013)  & (.016)  \\
Independence       & -.082***  & -.053***  & -.051**   & -.069*** &         &         & -.047**  &         &         \\
                   & (.015)    & (.014)    & (.016)    & (.017)   &         &         & (.014)   &         &         \\
Availability       & .090***   & .120***   &           &          &         & .187*** & .066*    &         & .186*** \\
                   & (.023)    & (.022)    &           &          &         & (.046)  & (.026)   &         & (.034)  \\
Closeness          & .062**    & .080***   & .069*     &          & .091**  &         & .054*    & .067**  & .065*   \\
                   & (.023)    & (.022)    & (.027)    &          & (.031)  &         & (.025)   & (.024)  & (.031)  \\
Likeability         & .138***   & .201***   &           & .178***  & .126*   &         & .166***  & .122**  &         \\
                   & (.030)    & (.029)    &           & (.053)   & (.053)  &         & (.037)   & (.041)  &         \\
Integrity          &           &           &           & .143**   & .109*   & .212*** & .108**   & .085*  & .097*  \\
                   &           &           &           & (.045)   & (.047)  & (.045)  & (.034)   & (.038)  & (.042)  \\
Reliability        &           &           & .085**    & -.078*   & .127**  &         &          & .119*** & .123**  \\
                   &           &           & (.031)    & (.039)   & (.039)  &         &          & (.033)  & (.040)  \\
Adopters           & .131***   & .149***   & .252***   &          &         &         &          &         &         \\
                   & (.021)    & (.021)    & (.023)    &          &         &         &          &         &         \\
Residence          &           &           & -.254***  &          &         &         &          &         &         \\
                   &           &           & (.060)    &          &         &         &          &         &         \\
Gender             &           &           &           &          &         &         & -.220**  &         &         \\
                   &           &           &           &          &         &         & (.081)   &         &         \\
Income             &           &           &           &          &         &         &          &         &         \\
                   &           &           &           &          &         &         &          &         &         \\
Constant           & -4.442*** & -4.903*** & -5.590*** & 1.378*** & .768*   & .159    & 1.005*** & .592*   & 0.049   \\
                   & (.276)    & (.281)    & (.345)    & (.308)   & (.320)  & (.388)  & (.235)   & (.231)  & (.273)  \\
\midrule
Observations       & 2738      & 2738      & 2738      & 986      & 1061    & 698     & 1481     & 1553    & 1008    \\
R$^{2}$         & .152      & .153      & .210      & .446     & .426    & .482    & .440     & .432    & .515    \\
Adj. R$^{2}$ &           &           &           & .442     & .422    & .478    & .437     & .429    & .512   \\
\bottomrule
\addlinespace[1ex]
\multicolumn{7}{l}{\textsuperscript{*}$p<.050$, \textsuperscript{**}$p<.010$, \textsuperscript{***}$p<.001$; Standard errors in parentheses}
\end{tabular}
 \label{tab:AttributeImportance}
\end{table}

\subsection{Peer group influence strength}
\label{sec:influence}

Respondents were asked to provide information on the impact of each peer contact they reported during each decision-making stage: first, they indicated whether each contact influenced their attitude or behaviour toward PV systems (positive influence, negative influence, no influence); for those contacts identified as positively or negatively influential, they were then asked to indicate the strength of influence using a 1-10 sliding scale (Section \ref{sec:communication}). Even though all social peer groups showed similar average influence ratings (see also Supplementary Material F), the peer group family and relatives was the most influential group:
\begin{itemize}
\item Family \& relatives: stage I (\textit{$7.639\pm2.025$, n=536}), stage II (\textit{$7.550\pm2.062$, n=538}), stage III (\textit{$7.734\pm2.055$, n=319});
\item Friends: stage I  (\textit{$7.350\pm1.980$, n=492}), stage II (\textit{$7.211\pm2.071$, n=533}), stage III (\textit{$7.252\pm2.148$, n=324});
\item Acquaintances \& co-workers: stage I (\textit{$7.312\pm2.058$, n=418}), stage II (\textit{$7.081\pm2.103$, n=455}), stage III (\textit{$7.387\pm2.180$, n=279});
\item Neighbours: stage I (\textit{$7.408\pm2.044$, n=391}), stage II (\textit{$7.185\pm2.162$, n=424}), stage III (\textit{$7.434\pm2.117$, n=260}). 
\end{itemize}
All peer groups showed a somewhat higher influence in stage I and stage III than in stage II, which suggests that residential decision-makers are less open to influence during this stage (the authors consider this plausible, as some decision-makers may be alternatively engaged in intensive self-study of the technology, but in any case, further research would be needed). As with peer attributes, respondents with a higher number of PV adopters in their social peer groups rated the influence of all peer groups as higher on average in all three stages than did respondents with a lower number of PV adopters in the social peer groups.

Lastly, if highly credible sources are indeed more influential than less credible sources, the perceived attributes of the different peer groups (Section \ref{sec:perception}) might give an explanation of the strength of the influence attributed to each reported peer contact. As with the contact initiation analysis, a linear step-wise regression was applied. An overview of the explanatory power and the statistically significant coefficients (\textit{p$<$.050}) of regression models for the different decision-making stages is outlined for the influences of self-initiated peer exchanges and for both self-initiated and peer-initiated exchanges in the middle panel and right panel of Table \ref{tab:AttributeImportance}. The $\beta-$coefficients allow an estimate as to what extent a change in a perceived attribute level impacts the strength of the contact influence of the peer group in each decision-making stage. The results demonstrate that the influence strength of the peer groups can largely be explained by their perceived attributes over all cases and in all stages (up to \textit{R\textsuperscript{2}= .512}). The control variables (gender, income level, place of residence, and number of peer PV adopters) did not significantly affect perceived influence strength of the peer groups.

Focusing on the actively initiated exchanges, we see that the most important attribute over all cases and stages at a significant level was competence (\textit{$\beta_{I}$=.320***; $\beta_{II}$=.347***; $\beta_{III}$=.257***}), followed by trustworthiness in stage I and stage III (\textit{$\beta_{I}$=.204***}; \textit{$\beta_{III}$=.214***}) as well as likeability in stage I and stage II (\textit{$\beta_{I}$=.178***}; \textit{$\beta_{II}$=.126*}). Attributes like integrity and power also had an effect on influence strength over all stages. At the same time, some attributes were of varying effectiveness in certain stages. This seems plausible, as decision-makers focus on different aspects during different stages; however, further research is required. Even though closeness played a subordinate role compared to the attributes mentioned above, a significant impact on influence strength during the second stage (\textit{$\beta_{II}$=.091**}) was identified. The same was found for availability during the last stage, which is plausible as this is the stage in which particular suppliers and service providers come into consideration (\textit{$\beta_{III}$=.187***}).

\section{Research implications}
\label{S:5}

\subsection{Results discussion}
PV adoption by social peers was found to have a clear impact on normative beliefs, as well as on perceptions of peer group credibility: respondents with numerous PV adopters in their peer groups were more likely to agree with statements on the normative benefits of adopting PV (Section \ref{sec:pressure}) and were more likely to rate their peers as competent, trustworthy, and available with regard to PV (Section \ref{sec:perception}). 

From one perspective, our findings confirm the prior research finding (and common-sense expectation) that residential decision-makers who believe in the benefits of adopting PV, and who have the practical capacity to adopt PV, tend to harbour stronger intentions to adopt PV. Korcaj et al.~\cite{Korcaj.2015} revealed that positive attitudes towards PV, stronger pro-PV subjective norms (behaviour of peers, expectations of peers), and higher behavioural control lead to a stronger intention to adopt. Within Korcaj et al.'s sample, social status drivers, autonomy, financial and environmental benefits and in general risk and cost explained more than two-third of PV attitudes. Jacksohn et al.~\cite{Jacksohn.2019} and O'Shaughnessy et al. \cite{OShaughnessy.2020} showed that household income (and, by extension, behavioural control) also has a strong influence on adoption behaviour, a finding that the regression results presented in Section \ref{sec:pressure} confirm. %High-income households adopted more likely than low- and medium-income households according to our data which was also shown by O'Shaughnessy et al. \cite{OShaughnessy.2020}. %Regional characteristics as highlighted in Baginski and Weber~\cite{Baginski.2019}, however, played a very subordinate role with respect to our results.

From another perspective, our results also lend strength to peer effect hypotheses, which posit a causal relationship between geographical proximity to and/or social contact with other PV adopters and the decision to adopt \cite{Bollinger.2012,Graziano.2015,Rode.2020}. Similarly to Rai et al.~\cite{Rai.2016b}, Mundaca et al. \cite{Mundaca.2020}, and Petrovich et al.~\cite{Petrovich.2019}, the study found that the presence of PV adopters in decision-makers' neighbourhoods and/or peer groups motivates them to consider PV more seriously. %Furthermore, respondents recognized the importance of neighbours as information pathways in their decision by agreeing or strongly agreeing that without PV systems in their neighbourhood they would not have chosen to install. 
The study improves upon the temporal resolution of the peer effect hypothesis by incorporating a stage model of residential PV decision-making: it was found that potential and current adopters' self-identified progress through sequential decision stages was driven by the number of PV adopters in their social peer groups. A complementary correlation was found between number of peer PV adopters and stated intention to adopt. 

The study also sheds light on the interactive dynamics at work in peer effects. Not only were respondents with a higher number of PV adopters in certain peer groups were more likely to have had PV-related contact with members of these groups during all three stages in the decision-making process, but this contact was more likely to be bidirectional and self-initiated (Section \ref{sec:communication}). Again, clear parallels can be drawn with the results of Rai et al. \cite{Rai.2016b} who state that the majority of their respondents strongly agreed that talking to other owners was useful. Furthermore, within Rai et al.'s sample, respondents who reported fewer PV systems in their neighbourhoods indicated that they would have liked to talk to more PV owners, but couldn't find any. Based on these findings and the findings of the present study, a conclusion can be drawn that the presence of PV systems within one's neighbourhood and/or social group can act as a stimulus to information-seeking behaviours that are both more active and more sustained. Indeed, respondents in the present study were more likely to actively initiate PV-related contact with social peers than to be drawn into peer-initiated contacts, a finding which echoes Palm \cite{Palm.2017}. Respondents with more peer adopters in their social circles were more likely to report having initiated contact with peers, as were respondents who had either adopted or strongly intended to adopt PV. Following M\"uller~\cite{Rode.2020}, geographically and/or socially proximate PV systems and adopters can be interpreted as a pool of information which reduces uncertainty throughout the PV decision-making process, expediting self-directed progress through its stages. %A reason could be that respondents without peer adopters in their circle rather doubt their ability to find all necessary information for an optimal decision \cite{Scheller.2020}. 
%In line with Rogers \cite{Rogers.2003}, our results showed that respondents who reported less technology consciousness and social benefits (when compared to novelty seeking and independent judgement-making), reported a more passive behaviour with respect to contact initiation. 

Finally, the study contributes to our understanding of the subjective dimension of active peer effects by demonstrating that not only the quantity of peer adopters, but also the perceived qualities of peers contribute to effect strength. Respondents who assessed their peers as credible with regard to PV were more likely to initiate PV-related contact with peers (Section \ref{sec:communication}), as well as to report having been influenced by peer exchanges (Section \ref{sec:influence}). The analysis was controlled for gender, income, area of residence, and number of peer PV adopters, which revealed an intriguing distinction: active initiations of peer exchanges were driven by both the quantity of peer adopters and the perceived qualities of peers, whereas the influence ratings of these exchanges were almost entirely explained by the perceived attributes of peers. This result not only supports, but also adds nuance to the findings of Bollinger and Gillingham~ \cite{Bollinger.2012}, Rai and Robinson~\cite{Rai.2013}, and Rode and M\"uller~\cite{Rode.2020} that peer adopters cause active peer effects, as well as the finding of Palm~\cite{Palm.2016,Palm.2017} that geographically proximate peers who are perceived only as neighbours rather than also as friends are rarely identified as influential. We can understand this result to mean that active peer effects are multidimensional, and that different dimensions depend upon different properties of peer networks. Actors in networks with a high PV diffusion rate are more likely to engage in exchanges about PV, but once the exchange has started, its influence on intentions and behaviour is largely determined by the perceived qualities of the interlocutors. 

The results also suggest that influential peers possess a mix of positive qualities, and that the relative contributions of individual qualities to this mix depend upon the decision-maker's stage in the decision-making process. For instance, while competence proved important in all three stages, the importance of likeability decreased across the three stages; trustworthiness and availability were most important during the final stage (Section \ref{sec:communication}). Closeness is not negligible in any of the decision-stages, nor are integrity and power. It follows that the strongest peer effects would result from contact with credible peers with different combinations of specific positive attributes during different stages in the decision-making process. This again concurs with past studies such as Mundaca and Samahita~\cite{Mundaca.2020}, Wolske et al. \cite{Wolske.2020}, and Scheller et al.~\cite{Scheller.2020,Scheller.2021}. A possible explanation for the changing importance of different attributes throughout the decision-making process, as well as for the stronger influence of peer groups during the first and last stages of this process, could be that “consumers employ different evaluative criteria in alternative stages of their decision-making process" \cite{Arts.2011}. %The closer a behaviour comes in time, the more specific and context-dependent, and the less abstract and general considerations become \cite{Arts.2011}. 
Such a shift in the kind of knowledge deemed important by decision-makers, from informal knowledge at the awareness stage to concrete, economic and technical knowledge needed at later stages, was also visible in former qualitative research \cite{Scheller.2020,Scheller.2021}. 

\subsection{Policy recommendations}
From a practical perspective, our insights suggest that the effects of active peer exchanges rely on perceived attributes of the peers, with a combination of expertise, competence, and social charisma leading to the most capacity for influence. It follows that to promote PV adoption in a municipality, policymakers need to put socially well-connected residents at the centre of attention. Self-organised and participatory solar initiatives and interest groups can provide crucial support by cultivating relationships of trust between current and potential adopters. From a community angle, residential adopters could be recruited to participate in information campaigns such as neighbourhood seminars. From an individual angle, incentive programs such as recruit-a-friend could be launched. A multi-level governance framework is crucial here: since active initiations of peer exchanges are dependent in part on spatial proximity, neighbourhood districts may represent the administrative level with the highest potential for policy interventions designed to create momentum through peer effects. 

\section{Conclusion}
\label{S:6}

This paper seeks to contribute to the literature on peer effects in residential PV adoption by exploring the underlying mechanisms of active peer-to-peer exchanges that mediate the impact of such effects. A non-experimental survey approach was taken in order to enable the quantification and comparison of peer-to-peer mechanisms within sub-groups of residential PV decision-makers in Germany, for instance potential as opposed to current adopters. A three-stage model of the PV decision-making process was furthermore employed to shed light on the way such mechanisms might change over the course of the adoption journey.

The findings of the survey among potential and current adopters support the importance of active peer effects (without discounting the potential importance of passive effects), as well as the importance of known and well-regarded peers as opposed to merely geographically proximate peers. Decision-makers with PV adopters within their social circles were more likely to see PV adoption as a positive social norm and more likely to agree that PV could benefit them. They were also more likely to assess their peers as credible with regard to PV. This is crucial, as decision-makers who assessed their peers as credible (competent and trustworthy) were more likely to actively initiate peer exchanges and more likely to be influenced by these exchanges, during all stages of the decision-making process. These results indicate that both the quantity and the perceived qualities of peer adopters drive residential PV adoption throughout a substantial part of the decision-making process.

An inherent limitation of survey-based research is that respondents may assess their own perceptions, intentions, and behaviours differently in different contexts\cite{Wolske.2020}. For instance, a tendency exists to assess one's own decisions positively once they have already been made \cite{Rai.2016b}. In the case of the present study, respondents were asked to describe interactions during past stages in the decision-making process as well as their current stage; in some cases, a significant amount of time may have passed and the interactions may not be remembered accurately. Another limitation of the present study is that social networks were investigated on the level of peer groups rather than the level of individual peers, as was done for instance by Palm~\cite{Palm.2017}. Thus, the relationship between peer attributes and peer effects can only be quantified on an aggregate level; analyses of the type outlined in Graziano and Gillingham \cite{Graziano.2015} and Rode and Müller \cite{Rode.2020} are not possible. Survey formats are also ill-suited to capturing granular information on the context of peer interactions, which socio-technical transition researchers such as Geels et al. \cite{Geels.2018} identify as important to peer effects.

Furthermore, the exclusion of early-stage potential adopters and PV rejectors from the sample limits the range of comparative analyses that can be performed, as well as the analysts' ability to account for self-selection bias (as late-stage potential adopters and current adopters are, by definition, more interested in PV). Finally, investigating multiple stages in the decision-making process meant that limited time could be dedicated to each stage; it follows that a detailed view of the effect of active exchanges on the final decision itself, as provided by Rai and Robinson \cite{Rai.2013}, was not possible.

The present analysis has been performed as an intermediate step for the parametrization of an agent-based model (cf. \citep{johanning2020modular, Scheller.2019,reichelt2021towards}) that aims to simulate residential PV adoption decisions in specific spatial-temporal contexts. Future research could build on these findings to further clarify the role of peers in the adoption process, which could in turn improve policymakers' ability to harness peer effects to support individual decision-making. In particular, future research could further investigate the processes by which peer influence operates. Further data could also be gathered in different countries to establish the generalisability of the findings across different populations or cultures. A more detailed investigation of the proposed peer and relational attributes, expertness, trustworthiness, power, likeability and closeness, could also be helpful.   

\section*{Supplementary material} 
The Supplementary Material (SM) consists, first, of the survey (SM A) and the presentation and comparison of the sample (SM B). Second, insights of the analyses are presented regarding the benefit statements (SM C), peer perception (SM D), contact initiations (SM E), and influence strength (SM F).

\section*{Declaration of competing interest} 
The authors declare that they have no known competing financial interests or personal relationships that could have appeared to influence the work reported in this paper.

\section*{Acknowledgement}
The authors wish to thank Sinus Markt- und Sozialforschung GmbH for the cooperation in conducting the survey. Fabian Scheller receives funding from the project SUSIC (Smart Utilities and Sustainable Infrastructure Change) with the project number 1722 0710. This study is financed by the Saxon State government out of the State budget approved by the Saxon State Parliament. Furthermore, Fabian Scheller also kindly acknowledges the financial support support of the European Union's Horizon 2020 research and innovation programme under the Marie Sklodowska-Curie grant agreement no. 713683 (COFUNDfellowsDTU).

\appendix


%% References
%%   \cite{key}      ==>>  [#]
%%   \cite[chap. 2]{key}==>>  [#, chap. 2]
%%   \cite{key}      ==>>  Author [#]

\newpage
\bibliographystyle{elsarticle-num}
\bibliography{ModelDescription.bib}






% \end{thebibliography}

\end{document}